\documentclass[conference]{IEEEtran}
\IEEEoverridecommandlockouts

\usepackage{tikz}
\usepackage{amsmath}

\usepackage{xspace}
\usepackage{graphicx}
\usepackage[caption=false,font=footnotesize]{subfig}
\usepackage{booktabs}
\usepackage{array}
\usepackage[utf8]{inputenc}
\usepackage{cite}
\usepackage{cleveref}
\usepackage[ruled, vlined, linesnumbered]{algorithm2e}
\usepackage{listings}
\usepackage{xcolor}
\usepackage{balance}
\usepackage{url}

\crefname{section}{§}{§§}
\Crefname{section}{§}{§§}

\newcommand{\self}{VeriNC\xspace}
\newcommand{\sysname}{VeriNC\xspace}
\newcommand{\numsystems}{nine\xspace}
\newcommand{\fisslock}{\textsc{FissLock}\xspace}
\newcommand{\parab}[1]{\noindent\textbf{#1}\xspace}

\makeatletter
\def\ps@arxivcopyright{%
  \def\@oddhead{}\def\@evenhead{}%
  \def\@oddfoot{\parbox[t]{\textwidth}{\scriptsize\copyright~2026 IEEE. Personal use of this material is permitted. Permission from IEEE must be obtained for all other uses, in any current or future media, including reprinting/republishing this material for advertising or promotional purposes, creating new collective works, for resale or redistribution to servers or lists, or reuse of any copyrighted component of this work in other works.}}%
  \def\@evenfoot{\@oddfoot}%
}
\makeatother

\begin{document}

\title{\self: Finding Design Risks of In-Network Computing Systems}

\author{
\IEEEauthorblockN{Tianyu Bai\IEEEauthorrefmark{1}\quad
Xiaoxi Zhang\IEEEauthorrefmark{2}\quad
Haoqing Wang\IEEEauthorrefmark{1}\quad
Ying Zhang\IEEEauthorrefmark{3}\quad
Wenfei Wu\IEEEauthorrefmark{1}\IEEEauthorrefmark{4}\thanks{Corresponding author: Wenfei Wu.}}
\IEEEauthorblockA{\IEEEauthorrefmark{1}Peking University\quad
\IEEEauthorrefmark{2}Sun Yat-sen University\quad
\IEEEauthorrefmark{3}Meta}
\IEEEauthorblockA{\IEEEauthorrefmark{4}Beijing Key Laboratory of Software and Hardware Cooperative Artificial Intelligence Systems}
\IEEEauthorblockA{\IEEEauthorrefmark{1}\{tianyubai, whqwq\}@stu.pku.edu.cn, wenfeiwu@pku.edu.cn\quad
\IEEEauthorrefmark{2}zhangxx89@mail.sysu.edu.cn\quad
\IEEEauthorrefmark{3}Zhangying@meta.com}
}

\maketitle
\thispagestyle{arxivcopyright}
\pagestyle{plain}

\begin{abstract}

The emergence of programmable switches has brought in-network computing (INC) into the spotlight in recent years. By offloading computation directly onto the data transmission process, INC improves network utilization, reduces latency to sub-RTT levels, saves link bandwidth, and maintains throughput. However, INC disrupts the transparency of traditional networks, forcing developers to consider network exceptions like packet loss and out-of-order. If not properly handled, these exceptions can lead to violations of application properties, such as cache consistency and lock exclusion. Usual testing cannot exhaustively cover these exceptions, raising doubts about the correctness of INC systems and hindering their deployment in the industry.

This paper presents \self\footnote{\self is open-sourced at \url{https://github.com/cs-icez/verinc-compiler}.}, the first general-purpose tool for verifying INC systems. \self provides a high-level specification language and saves developers 67.2\% lines of code on average. To help better understand the behavior of the system, \self offers configurable network environments.
\self enables developers to express INC-specific correctness properties.
\self translates developer-specified systems into state transition representations, performs model checking to detect potential design risks, and reports violation traces to developers.
We propose optimizations for INC-specific scenarios to address the challenge of state space explosion. We modeled INC systems across four application domains and identified design risks with \self in seconds. \self has also been adopted to guide the design of a new INC protocol. Based on our verification experience, we summarize lessons that help develop a correct INC protocol.
We further reproduce them in real systems to confirm the validity of our verification result\footnote{Reproduction code is open-sourced at \url{https://github.com/cs-icez/verinc-violation}.}.

\textit{This work does not raise any ethical issues.}

\end{abstract}

\section{Introduction}
\label{sec:introduction}

The advent of programmable switches, including Barefoot To\-fi\-no~\cite{Tofino},  Cisco Silicon One~\cite{Cisco}, Broadcom Trident~\cite{Trident}, Juniper Networks Trio~\cite{Trio}, and Huawei NetEngine~\cite{NetEngine}, has opened up new possibilities for in-network computing (INC). These switches enable flexible user-defined packet processing at line speed and encourage the application developer to offload computation functions directly onto the data transmission process. Using programmable switches, INC applications can optimize network utilization, achieve sub-RTT latencies, conserve link bandwidth, and maintain high throughput.

Numerous INC applications have been proposed in recent years, including tensor aggregation~\cite{SwitchML, ATP, DSA, NetReduce}, key-value store cache~\cite{NetCache, DistCache, FarReach, Seer}, lock manager~\cite{NetLock, FISSLOCK}, virtual address translation~\cite{SwitchV2P}, task scheduling~\cite{RackSched, Horus} and consensus~\cite{P4xos, NetPaxos, NOPaxos}. For example, ATP~\cite{ATP} accelerates distributed ML training throughput by 38\%\textasciitilde 66\%
in a cluster shared by multiple jobs; Seer~\cite{Seer} achieves up to 65\% lower cache miss ratio and up to 78\% lower flow completion time compared to LRU for key network applications; \fisslock~\cite{FISSLOCK} cuts up to 79.1\% of median lock grant time in the microbenchmark and improves transaction throughput for TPC-C by 2.28$\times$. These applications demonstrate the potential of INC in various fields.

Guaranteeing \emph{system correctness} is essential yet particularly difficult for INC systems. An INC system, which manages both communication and computation for applications, must adhere to correctness criteria in terms of communication (e.g., transmission reliability) and computation (e.g., cache consistency and lock mutual exclusion). However, INC differs from host-based computation in two ways that make correctness uniquely difficult.

$\bullet$ \textbf{Network unreliability can lead to computational errors.} Networks may lose packets due to corruption or congestion. Conventional networks rely on TCP to ensure immutable byte-stream delivery, where hosts acknowledge packets and retransmit lost ones. However, such reliability is guaranteed only at endpoints and is not extended to intermediate network devices. In an INC system, even with TCP, the switch may still observe packet loss, duplication, and out-of-order. If not properly addressed, the switch will handle the altered stream in its usual manner. Such operations can further cause inconsistency in system states~\cite{FarReach}, loss of data~\cite{FISSLOCK}, or even preventing system termination (Section~\ref{ssec:tensor-aggregation-results}).

$\bullet$ \textbf{The limited programmability of switches complicates error recovery.} Programmable switches offer re\-strict\-ed programming capabilities, such as the absence of loops and limited access to switch memory (a single read/write per pipeline stage)~\cite{P4-Survey}. Also, the switch data plane cannot support reliable communication~\cite{RedPlane}. Developers have to significantly simplify the protocol in order to fit it into the switch. Thus, when a network exception occurs, the switch might lack the necessary state information or logical processing ability to resolve the issue, leading to potential errors. This fundamentally distinguishes INC systems from traditional distributed systems, where nodes are general-purpose servers with rich computation and storage resources.

Due to the challenges in ensuring system correctness, INC systems often encounter doubts about their suitability for deployment in production environments.

\textbf{INC developers need a tool for comprehensive exploration of network exceptions.} Studying existing literature~\cite{FISSLOCK, ATP, NetReduce, SwitchML, NetCache, FarReach, NetLock}, we find that developers tend to propose algorithms designed to work correctly in reliable networks and subsequently handle exceptions with patch mechanisms. However, people are not adept at identifying corner cases; thus, relying on human intuition does not ensure all-case correctness due to (1) unforeseen exceptions not being considered, (2) patch mechanisms addressing only specific exceptions, rather than the underlying root causes, and (3) potential conflicts between different patch mechanisms, especially in extreme situations where multiple exceptions co-occur.
For instance, our experiments demonstrate that under particular packet loss and event interleaving, NetCache~\cite{NetCache} and FarReach~\cite{FarReach} fail to guarantee the consistency of the cache system (Section~\ref{ssec:key-value-results}).
Developers generally use synthetic workloads~\cite{YCSB} or simulators~\cite{ns3} to test systems, but these methods are insufficient for capturing all exceptional cases.

\textbf{Current network verification tools struggle to satisfy the requirements of INC.} (1) Traditional networks, which do not involve computation, have led prior research to focus mainly on routing properties, such as reachability analysis~\cite{Reachability1, Reachability2}, loop detection~\cite{HSA, VeriFlow}, and isolation verification~\cite{Isolation1, AP-Verifier}. In contrast, we aim to verify application properties involving both communication and computation. (2) Since traditional packet stream is an end-to-end concept, many studies consider stateless networks~\cite{HSA} or focus on one-packet-at-a-time processing~\cite{NetSMC}. Conversely, INC protocols necessitate switches to maintain state information and correctly process multiple interrelated packets. (3) Many existing studies prioritize verifying implementation correctness over design correctness: they prove that implementation code aligns with design specifications~\cite{DNS-V}, rather than that the design itself satisfies correctness properties. Notably, recent efforts have aimed at verifying P4\footnote{P4~\cite{P4} is a language for programming the data plane of network devices.} programs~\cite{p4v, Vera, ASSERT-P4, Aquila}. They only verify properties concerning the data plane of programmable switches and thus do not align with our objectives. %

\textbf{We present \self, the first general-purpose INC system verification tool.} %
Developers use the \self specification language to write system specifications and desired properties, which are then checked by \self. If the verification succeeds, \self simply returns success. If the verification fails, \self returns a violation trace, which can help developers gain deeper insights into the system's behavior.

First, \self provides a high-level abstraction model and specification language tailored to INC systems. In the abstraction of \self, a system is composed of nodes, each running several threads simultaneously, connected through links. Threads of the same node synchronize with private variables and communicate with other nodes by exchanging packets. \self directly provides reliable, lossy, and out-of-order network environments, together with primitives for packet transmission and thread execution, so users can specify protocol behavior under different network conditions without modeling the network mechanisms themselves.

Second, \self uses linear temporal logic (LTL) formulas to express correctness properties and utilizes model checking to explore all possible execution traces of concurrency units. The \self compiler translates the \self specification into a lower-level state transition representation, allowing us to leverage the highly optimized performance of state-of-the-art model checkers~\cite{SPIN,tlaplus}.

Finally, we propose optimizations tailored to INC scenarios to address the challenge of state space explosion in model checking, including input space reduction, symmetric state elimination, and network nondeterminism constraints. These optimizations allow the state space to be fully explored within minutes on a reasonable parameter size.

We have used \self to model \numsystems INC systems, spanning tensor aggregation~\cite{SwitchML, ATP, NetReduce, SwitchGNN}, key-value caching~\cite{NetCache, FarReach}, lock management~\cite{NetLock, FISSLOCK}, and collective communication~\cite{EPIC}. We found design risks in several of these systems under certain network environments, including violations of terminality and application properties. These violations involve many interrelated packets and long logical chains, making them difficult to identify through human intuition or usual testing. %
To confirm the validity of our verification result, we further reproduce these violations with real implementations.

In summary, we make the following key contributions:

\begin{itemize}
\item We propose \self, the first general-purpose INC system verification tool, allowing users to model INC systems and verify correctness properties.
\item We introduce optimizations tailored to INC scenarios to improve verification efficiency.
\item We apply \self to model \numsystems INC systems spanning several application domains, identifying design risks, proposing fixes, reproducing the violations with real implementations, summarizing lessons learned, and guiding new protocol design.%
\end{itemize}

\section{Background and Motivation}
\label{sec:motivation}

\subsection{Model Checking}
\label{ssec:model-checking}

\parab{Preliminaries.}
In model checking, a user delineates the system as a \emph{model} with a set of \emph{variables}. The values of all variables define the \emph{global state} of the system. Each operation that modifies the global state defines a state \emph{transition}. All global states reachable from an initial state, together with the possible transitions, form a directed graph called the \textit{state space}. Each path within this graph, from the initial to a terminal state, signifies an \textit{execution trace}. A system property may be a constraint on an individual state, depicted with first-order logic (FOL), or a relationship between states along an execution trace, depicted with linear temporal logic (LTL). Model checking~\cite{MC1} provides a search algorithm to verify these formulas with a given model.

\parab{Workflow.}
The model checking algorithm performs \textit{state space exploration} alongside \textit{property violation detection} to verify the system model. It maintains a directed state graph $G$, and a queue $Q$ for state frontiers, traversing the state space in a breadth-first manner. The model checker
starts by inserting all initial states into $G$ and $Q$. During each iteration, it removes the first state $s$ from $Q$ and determines the set $T$ of its successor states. If $T$ is empty, it reports a deadlock and halts. Alternatively, for each state $t$ in $T$, it evaluates whether the transition $s \to t$ violates any properties. If a violation occurs, it reports an error and halts. Otherwise, it adds the edge $s \to t$ to $G$. If $t$ is not be present in $G$, it appends $t$ to both $G$ and the end of $Q$. Special considerations are taken for termination states of the system. Once $Q$ is empty, the model checker halts and declares the verification successful.

\parab{Why model checking?}
In addition to model checking, there are other formal verification methods, such as theorem proving~\cite{Theorem-Proving} and SMT solvers~\cite{SMT}. We chose model checking because network protocols naturally map to state machines. As a result, model checking is easily accepted by network practitioners, and it is straightforward to model networked systems using state-transition representations. In contrast, theorem proving and SMT solvers require extracting the logical structure of the system, which is highly non-trivial. Furthermore, it is challenging for theorem proving to achieve automated proofs and counterexample generation~\cite{COMST18}.

\subsection{Model Checking for INC}
\label{ssec:model-checking-inc}

Modeling the computational behaviors of systems has been extensively researched~\cite{verify-paxos,verify-other}, but model checking specific to INC introduces new requirements for the network part.

\parab{Providing friendly network abstractions in the specification language.} An INC system spans almost every layer of the network stack (device, routing, transmission, and application), making it difficult and unnecessary to model all the network details. For example, the switch hardware programming language (e.g., P4) provides low-level hardware instructions; translating them into high-level language is cumbersome and error-prone. The network reliability mechanism includes packet acknowledgment, windowing, and retransmission; describing the overall behavior is as complex as developing the transmission layer. Furthermore, some low-level operations are well-studied and validated over long-term network use, thus not needed to verify its internal correctness. \self provides interfaces for users to specify network topology, routing, and transmission.

\parab{Analyzing network nondeterminism in state space exploration.} Even for nodes directly connected via a link, packet loss may still occur due to buffer overflow or bit corruption. Packet loss further leads to retransmissions that may result in duplicated or out-of-order packets. This type of network nondeterminism significantly contributes to errors in INC systems. Although programming languages abstract away transmission and hardware details, model checking must incorporate this nondeterminism. Doing so enables a comprehensive exploration of the state space, aiding in detecting errors in corner cases. \self offers three types of network environments: reliable, lossy, and out-of-order. They are encapsulated in the primitives (\texttt{Send}, \texttt{Multicast}, and \texttt{Receive}) \self provides.

\parab{Specifying correctness properties of both communication and computation.} The INC system orchestrates both communication and computation processes; thus, it must guarantee both correctness. Network-specific properties include protocol terminality and memory leak freedom in switches. Computation-specific properties include application-specific correctness properties (Section~\ref{sec:use-case}).

\subsection{Challenge of State Space Explosion}

\parab{Model checking suffers from severe state space explosion problems.} Data-intensive INC systems handle a collection of items and involve multiple endpoints, leading to an expansion of states and transitions. Network nondeterminism results in multiple potential outcomes per packet sending, further exacerbating the expansion. By applying INC domain insights, such as data independence, node symmetry, and impact of network nondeterminism, we introduce optimizations of input space reduction, symmetric state elimination, and network nondeterminism constraints, effectively minimizing the state space and improving verification efficiency (Section~\ref{sec:optimization}).

\section{\self Design}
\label{sec:design}

\subsection{Overview}
\label{ssec:overview}

\begin{figure}[!ht]
\centering
\includegraphics[width=0.47\textwidth]{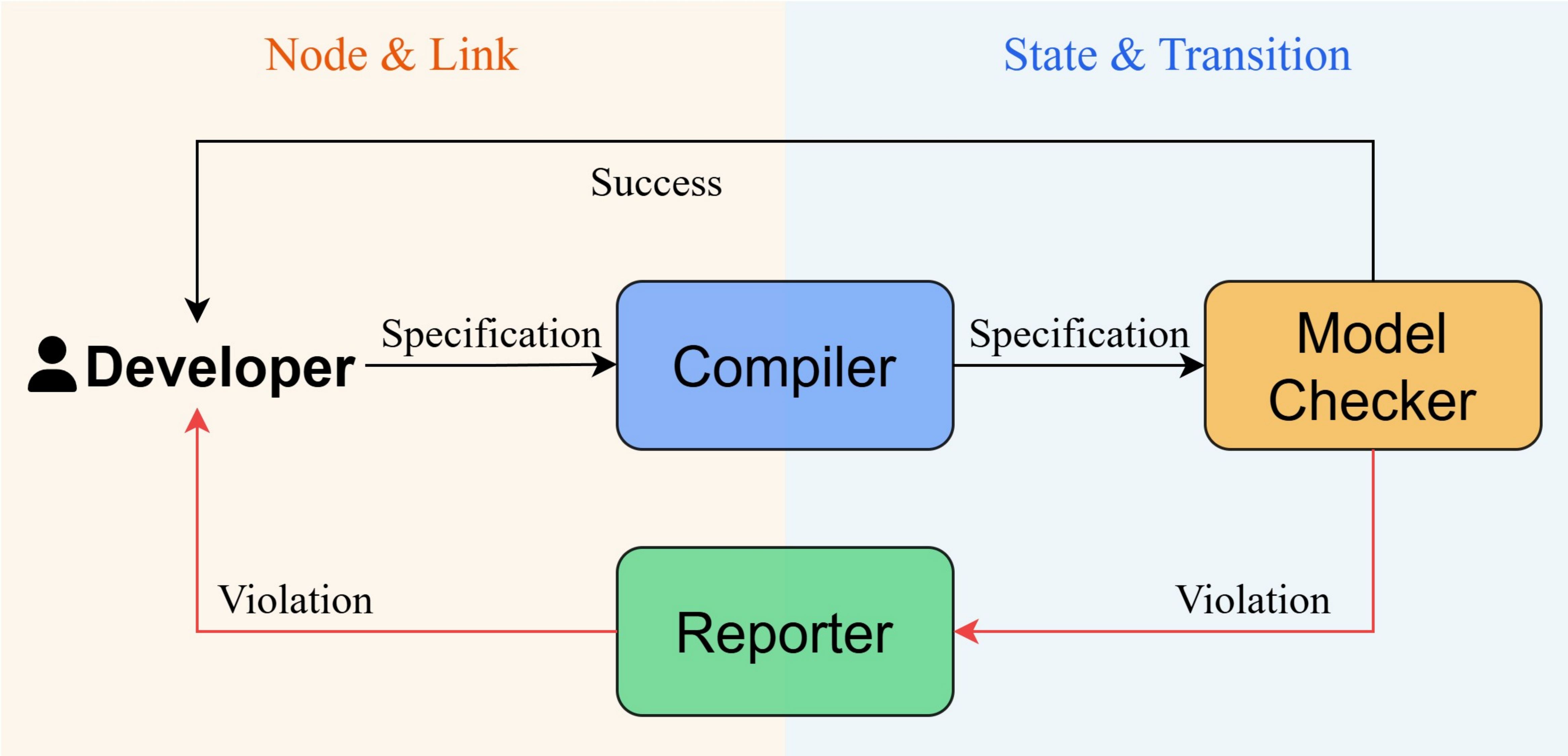}
\caption{The workflow of \self. The left half deals with nodes and links, while the right deals with states and transitions.}
\label{fig:workflow}
\end{figure}

Figure~\ref{fig:workflow} demonstrates the structure and workflow of \self. It consists of a specification language, a compiler, a model checker, and a reporter. \self provides a high-level specification language featuring network abstractions, such as topology and transmission, enabling users to create a network, specify the INC protocol, and express correctness properties.

The user submits its INC system specification to the \sysname compiler, which converts it into a low-level state transition representation. The model checker then reads the converted specification and performs verification, either declaring success or outputting a violation trace. The reporter interprets this trace, returning it to the network abstractions for user review.

\subsection{INC System Abstraction}
\label{ssec:system-model}

The abstraction model of \self consists of nodes and their networking.

\parab{Node.} Each node corresponds to a machine, with multiple threads assigned to different tasks. For example, the application thread executes the application logic, the receive thread monitors the receive buffer for incoming packets, and the retransmission thread initiates retransmissions upon timeout. Nodes have private variables for state recording and intra-node synchronization.

Threads run concurrently. An interleaved execution selects a thread, interrupts it at any moment, switches to another non-blocked thread, and continues until system termination. Model checking explores every potential interleaving. Threads can invoke \texttt{Wait(condition)} to block itself until the condition is satisfied. A blocked thread will not be selected for execution.

Threads terminate after executing their code or can invoke \texttt{Exit} to terminate the entire node. The system terminates once all threads terminate.

\parab{Network.}
Nodes are interconnected via links, creating a network structure. Inter-node communication involves packet exchange. Nodes generate packets and trigger \texttt{Send}. The next-hop node retrieves the packet by invoking \texttt{Receive} and decides to either discard, respond, alter, or forward it. In \self, packets are immediately directed to the next hop when sent. As model checking explores all possible interleaving, the receiver might process the packet at any moment, effectively simulating packet delays. Likewise, transmission timeout can be triggered at any time.

\parab{Nondeterministic network environment.}
Real-world networks exhibit unreliability. \self offers three types of network environments: reliable, lossy, and out-of-order. A reliable network ensures all packets arrive sequentially at the next hop. Conversely, in a lossy network, packets may not arrive. In an out-of-order network, packets sent subsequently may precede earlier ones at the next hop. A network can be both lossy and out-of-order.

\subsection{Specification Language}
\label{ssec:language}

\self defines a specification language allowing users to model an INC protocol conveniently. A specification comprises four components: configuration, topology, protocol, and property. At the top level, the configuration block defines parameters of the verification process; the topology block declares node types, nodes, links, and routes; the protocol block defines variables and threads; and the property block specifies the correctness properties to be verified. The formal grammar is provided in Appendix~\ref{sec:grammar}.

\parab{Configuration.} In \self, customizable parameters include the network environment, its nondeterminism constraints, and %
user-defined constants. Users can designate the network as reliable, lossy, or out-of-order. Finer-grained configuration is available, e.g., setting specific links to be lossy while others remain reliable. Moreover, network nondeterminism is controllable (Section~\ref{sec:optimization:extremity}). Users can define other constants like the number of requests.

\parab{Topology.} In \self, users interact with nodes and links directly, where every node is assigned a type, like client or switch. Nodes of the same type behave identically. Users define node types, instantiate nodes, and establish links among them. Users may manually specify routing tables for each node, or leave part of the work to the compiler.

\parab{Protocol.} An INC protocol essentially defines the states and behaviors of each type of node. \self specifies node states as variables, supporting integers, strings, sequences, sets, and dictionaries. \self specifies node behaviors as threads. Like in usual programming languages, a thread consists of statements, either simple or compound. Simple statements include breakpoints, expressions, assignments, and temporary value definitions, while compound statements include branching and looping. Expressions include binary operators, function calls, and first-order predicates for conditioning.

$\bullet$ \textit{Network-specific language element.}  \self provides the following abstractions for network transmission: (1) A packet is a dictionary with field names as keys. (2) \texttt{Send()} and \texttt{Receive()} describe the single-packet transmission between nodes. (3) \texttt{Multicast()} (along with \texttt{Receive()}) describes the process where the source replicates a packet and sends it to multiple destinations.

$\bullet$ \textit{Breakpoint.} Interpreting every statement as a transition would lead to an overwhelming expansion of the state space. \self runs model checking in a coarser granularity by requiring users to explicitly designate breakpoints in threads. Thread suspension and switch only occur at breakpoints. \texttt{Wait}, \text{Exit}, \texttt{Send}, \texttt{Multicast}, and \texttt{Receive} are mandatory breakpoints. To expose all intermediate states, each variable may be modified at most once between two adjacent breakpoints, enforced by the compiler. Users resolve a flagged duplicate update with either a \texttt{temp} statement or an additional breakpoint (Appendix~\ref{sec:compilation}). \texttt{temp} statements define temporary values, which are only valid between two adjacent breakpoints and thus not part of the system state.

$\bullet$ \textit{Modeling failure.} \self does not explicitly model component failure. Yet, by configuring a specific link to always drop packets, link failure is effectively simulated. Node failure can be simulated by implementing the corresponding logic manually, e.g., pausing all threads for some period and resuming with a certain state and an empty buffer.

An example thread is provided in Appendix~\ref{sec:example-thread}.

\parab{Property.} \self offers three methods for expressing correctness properties. The first is statement \texttt{Assert(expr, error\_msg)}, where \texttt{expr} is a first-order logic (FOL) formula that refers to the node's private variables and is verified when the statement is executed.

The second method is to define them as linear temporal logic (LTL) formulas within the property block, which can involve any variables in the system and are verified after each state transition. \self supports the LTL temporal operators \texttt{[]} (always) and \texttt{<>} (eventually), using TLA+ notation. We elaborate on the correctness properties for each INC system in Section~\ref{sec:use-case}.

The third method is to use a custom stateful check function backed by global variables for complex temporal properties. These global variables are virtual: they are referenced only by the check function and do not affect the protocol logic. In our evaluation, only cache coherence requires such a function (Section~\ref{ssec:key-value-results} and Appendix~\ref{sec:cache-consistency}).

\begin{figure}[!ht]
\centering

\begingroup
\definecolor{comment}{rgb}{0, 0.5, 0}
\definecolor{type}{rgb}{0.17, 0.57, 0.69}
\definecolor{number}{rgb}{0, 0.5, 0}

\lstset{
    backgroundcolor=\color{white},   %
    frame=single,                    %
    rulecolor=\color{black},         %
    basicstyle=\ttfamily\small,      %
    numbers=left,                    %
    numberstyle=\tiny\color{gray},   %
    numbersep=5pt,
    xleftmargin=2em,
    breaklines=true,                 %
    commentstyle=\color{comment},    %
    comment=[l]{//},
    keywordstyle=[1]\color{blue},    %
    keywordstyle=[2]\color{type},
    keywords=[1]{configuration, topology, protocol, nodetype, link, var, thread, null, route},
    morekeywords=[2]{Client, Switch, Controller, Server},
    literate=
        {c1}{c1}2
        {c2}{c2}2
        {c3}{c3}3
        {0}{{{\color{number}0}}}1
        {1}{{{\color{number}1}}}1
        {2}{{{\color{number}2}}}1
        {3}{{{\color{number}3}}}1
        {4}{{{\color{number}4}}}1
        {5}{{{\color{number}5}}}1
        {6}{{{\color{number}6}}}1
        {7}{{{\color{number}7}}}1
        {8}{{{\color{number}8}}}1
        {9}{{{\color{number}9}}}1
}

\begin{lstlisting}
configuration {
  MAX_LOSS = 0;
  MAX_OUT_OF_ORDER = 0;
  DUPLICATION = 0;
  TERMINATION_CHECK = 1;
  CLIENT_NUM = 3;
  REQ_NUM = 5;
  ...
}
topology {
  nodetype Client, Switch, Controller, Server;
  Client c1, c2, c3;
  Switch sw; Controller ctrl; Server s;
  link c1, c2, c3 -- sw -- ctrl, s;
  link ctrl -- s;
  route(c1, c2, c3) { s: sw; }
  route(sw) { ... }
  route(ctrl) { ... }
  route(s) {
    c1, c2, c3, sw: sw;
    ctrl: ctrl;
  }
}
protocol {
  var(Client) base = 1, requests = ..., replies = ...;
  var(Switch) cached = 0, valid = 0, value = null;
  ...
  thread(Client) ClientApp { ... }
  thread(Client) ClientRecv { ... }
  thread(Client) ClientRetx { ... }
  thread(Switch) SwitchRecv { ... }
  ...
}
// Properties are checked with Assert in threads.
\end{lstlisting}
\endgroup

\caption{The specification of NetCache.}
\label{fig:netcache-spec}
\end{figure}

Figure~\ref{fig:netcache-spec} presents our NetCache~\cite{NetCache} specification under a reliable network (no packet loss or out-of-order), prohibiting duplication (Section~\ref{sec:optimization:extremity}), enabling terminality check (Section~\ref{ssec:tensor-aggregation-results}), and involving three clients initiating five requests each. The network comprises nodes of four types, connected by links (\texttt{-}\texttt{-} denotes a full connection between operands), routing tables partially specified. Each type of node is associated with variables and threads. There is no property block as properties are checked with \texttt{Assert}s.

\sysname compiles specifications to TLA+~\cite{tlaplus} for model checking; the reporter translates violation traces back to network-level abstractions for user review. By replacing the compiler backend and reporter frontend, it is easy to switch to another model checker. Compilation details are provided in Appendix~\ref{sec:compilation}.

\section{Optimization}
\label{sec:optimization}

\subsection{Reducing Input Space}
\label{sec:optimization:item}

Model checking is mainly appropriate for control-intensive systems and less suited for data-intensive ones, as data typically ranges over infinite domains~\cite{MC-Principle}. INC systems are driven by requests and suffer a similar problem. We exploit INC characteristics to reduce the input space to a manageable size.

INC applications often handle logically independent objects, and it is straightforward to ensure independent access by design. We assume this as a precondition, concentrating on states concerning a specific object, such as a task, a key-value pair, or a lock.

INC protocols maintain only very few state variables in the switch per object due to the limitation of programmable switches. When processing long request sequences, the same state will occur repeatedly. Our findings suggest that a relatively small number of requests suffice to identify inconsistent states within the system. Consequently, we limit the input request sequence to a short length in verification.

\subsection{Leveraging Node Symmetry}

An INC system consists of multiple nodes, but of only few node types. Nodes of the same type exhibit symmetry. For example, when client 1 issues a read request to the storage server and client 2 issues a write, it is essentially the same as client 1 issuing a write and client 2 issuing a read.

For a system state, swapping all variables associated with two nodes of the same type constitutes a swap of the state. The composition of swaps forms a permutation. In model checking, any new state is discarded if one of its permutations has been explored. This optimization, which is called symmetry reduction, maintains the completeness of verification~\cite{MC-Symmetry}.

\subsection{Constraining Network Nondeterminism}
\label{sec:optimization:extremity}

Unreliable networks and retransmissions impose a considerable burden on model checking. Each packet may be discarded in a lossy network; each packet could be inserted at any position in the receiver’s buffer in an out-of-order network; retransmissions might occur at any time. In the worst scenario, where all packets are lost, the system never terminates.

We provide users with configurable parameters to constrain the network nondeterminism, including maximum packet losses, unreliable links, out-of-order packets, and duplicates. These bounds can be configured either globally across all links or separately for each link. A packet that is retransmitted when still present in the network is considered a duplicate. In our findings, only a small amount of network nondeterminism is sufficient to expose risks in INC systems.

\section{Discussion}

\parab{Off-the-shelf model checker.} \self uses an off-the-shelf model checker, TLA+, rather than implementing one from scratch. A mature model checker avoids internal bugs and lets us leverage its optimized performance, but limits further optimizations to features TLA+ provides. We plan to replace TLA+ with our own custom model checker, validate its execution against TLA+ to ensure correctness, and explore optimizations at the algorithmic level based on the characteristics of INC systems.

\parab{Comparison with general-purpose specification languages.} Since \self compiles to TLA+, a natural question is what \self offers beyond writing TLA+ specifications directly. \self does not claim novelty at the level of verification algorithms. Its value lies in encapsulating INC-specific modeling patterns: configurable per-link network environments are provided as built-in primitives, freeing users from manually encoding network nondeterminism; optimizations such as symmetry reduction and network nondeterminism constraints are applied automatically; and violation traces are translated back to network-level abstractions, enabling developers to diagnose design risks without understanding TLA+ internals.

\parab{Applicability.}
In terms of language expressiveness, \self can model any INC system whose switches and end hosts coordinate through message passing. In practice, exhaustive checking is best suited to systems with relatively simple topologies and moderate node counts. Complete host-side protocol stacks beyond the INC logic, such as a full TCP stack, are outside our scope.

\parab{Manual specification writing.} \self requires users to manually write the protocol specification, rather than generating from implementation code (such as P4 and C++). This is an intentional choice. Our goal is to help users model the core logic, verify its correctness, and continuously refine the protocol at design phase, without introducing too much implementation details. This aligns with the fundamental purpose of model checking~\cite{Coherence-TLA}. The guarantees established by \self apply to the user-written model, not to its implementation.

\parab{Quantifying robustness to network exceptions.}
\self can quantify the degree of network nondeterminism that a protocol tolerates. By varying network nondeterminism constraints (Section~\ref{sec:optimization:extremity}), users can determine the largest checked bounds under which a property holds and the smallest bounds that produce a counterexample.

\parab{Hardware-aware abstractions.} Although we do not wish to introduce implementation details, it is important to note that the execution model of programmable hardware differs from that of end hosts, where design risks may stem. Currently, \self abstracts only general computation and provides no built-in abstractions for hardware-specific features such as packet-processing pipelines or TCAM storage. We leave hardware-specific abstractions to future work.

\section{Verification Performance}
\label{sec:evaluation}

\begin{table}[tb]
\centering
\caption{LOC and time to first violation of seven of the studied systems. The remaining two systems (SwitchGNN and EPIC) are discussed in \S\ref{sec:use-case}.}
\label{tab:LOC}

\small

\begin{tabular}{c c c c}

\toprule
\textbf{System} & \textbf{\self LOC} & \textbf{TLA+ LOC} & \textbf{Time (s)} \\
\midrule
SwitchML & 98 & 440 & 3 \\
ATP & 170 & 515 & {2} \\
NetReduce & 146 & 486 & {72} \\
NetCache & 190 & 589 & 8 \\
FarReach & 246 & 640 & 4 \\
NetLock & 148 & 504 & 4 \\
\fisslock & 466 & 1053 & 1 \\
\bottomrule

\end{tabular}
\end{table}

\parab{Implementation and experiment settings.}
We prototyped \self, including the compiler and the reporter. We used TLA+~\cite{tlaplus} as the model checker. The compiler is built on Flex/Bison~\cite{flex-bison} with around 3500 lines of C++, converting \self specifications into a TLA+ compatible format.%

The experiments were carried out on a workstation with a 4-core 2.2 GHz AMD EPYC-Milan processor, 4 GB DRAM, and 40 GB SSD.
We present evaluations of LOC, verification efficiency, the effect of optimizations, and the overhead of property checking.

\parab{LOC.} We modeled \numsystems state-of-the-art INC systems. Table~\ref{tab:LOC} shows the lines of code required to model each system using the \self specification language, as well as the lines of TLA+\footnote{Technically, PlusCal~\cite{PlusCal}, which is an intermediate language specified by TLA+.} code generated by the \self compiler. On average, \self reduces code by 67.2\%. Since we only modeled the core logic described in the paper (thus verifying design correctness rather than implementation), the LOC required for modeling are much fewer than their actual P4 implementation.

\parab{Verification efficiency.} Table~\ref{tab:LOC} presents the time to identify the first violation for each system, with parameters properly set. %
\sysname requires only a few to tens of seconds to detect violations. These violations are discussed in Section~\ref{sec:use-case}. INC characteristics make violations likely in shallow layers of the state space, so breadth-first search finds the first violation quickly.

\begin{figure}[!ht]
\centering

\subfloat[Time and diameter.]
{
\label{fig:efficiency-1}
\includegraphics[width=0.45\linewidth]{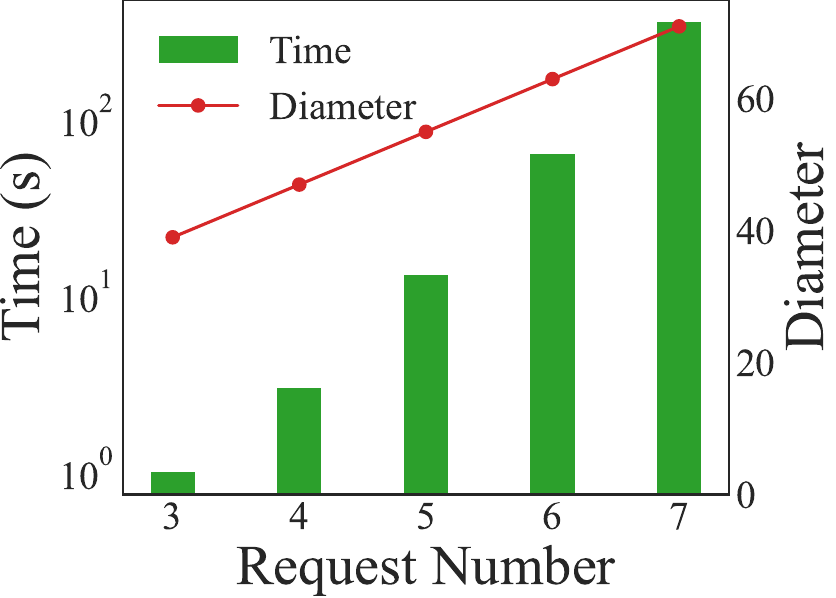}
}
\subfloat[State space size.]
{
\label{fig:efficiency-2}
\includegraphics[width=0.45\linewidth]{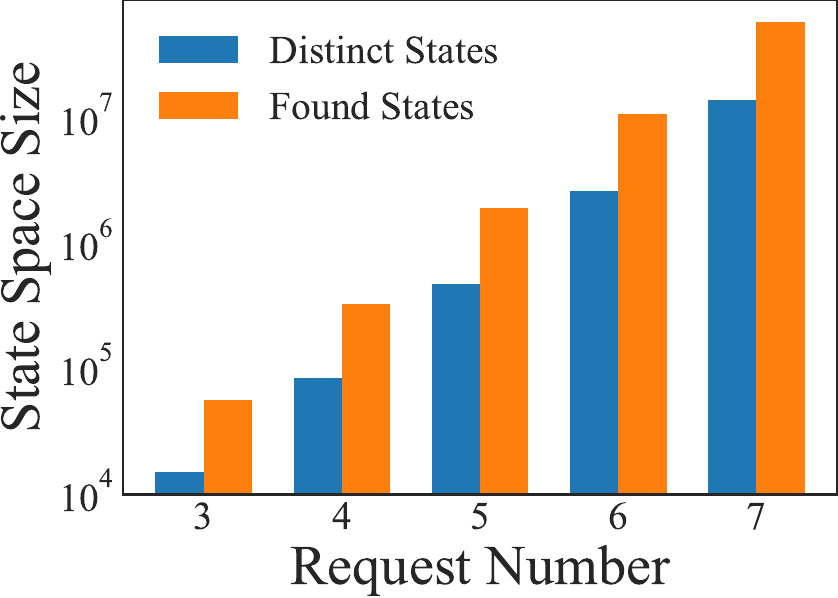}
}

\caption{The verification time, state space diameter, and state space size of NetCache.}
\label{fig:efficiency}
\end{figure}

Figure~\ref{fig:efficiency} shows the full state space size of NetCache under a reliable network with a single client. %
The diameter of the state space refers to the maximum distance from the initial state to all reachable states. The number of found states is the total explored, while distinct states is the de-duplicated count. As shown, under reasonable parameter settings, the full state space can be explored in a few seconds to a few minutes.

When no violation is found, \self explores the entire state space. When multiple violations coexist, \self reports the shallowest one with breadth-first search; users fix it and rerun verification to expose subsequent violations.

\begin{figure*}[!ht]
    \begin{minipage}[b]{0.39\textwidth}

        \centering

        \subfloat[Verification time.]
        {
        \label{fig:symmetry-1}
        \includegraphics[width=0.44\linewidth]{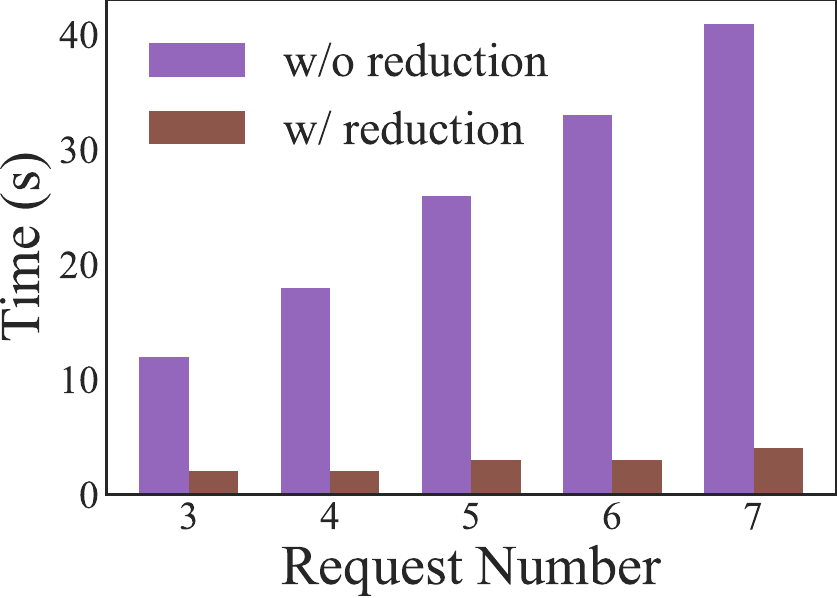}
        }
        \subfloat[Distinct States.]
        {
        \label{fig:symmetry-2}
        \includegraphics[width=0.44\linewidth]{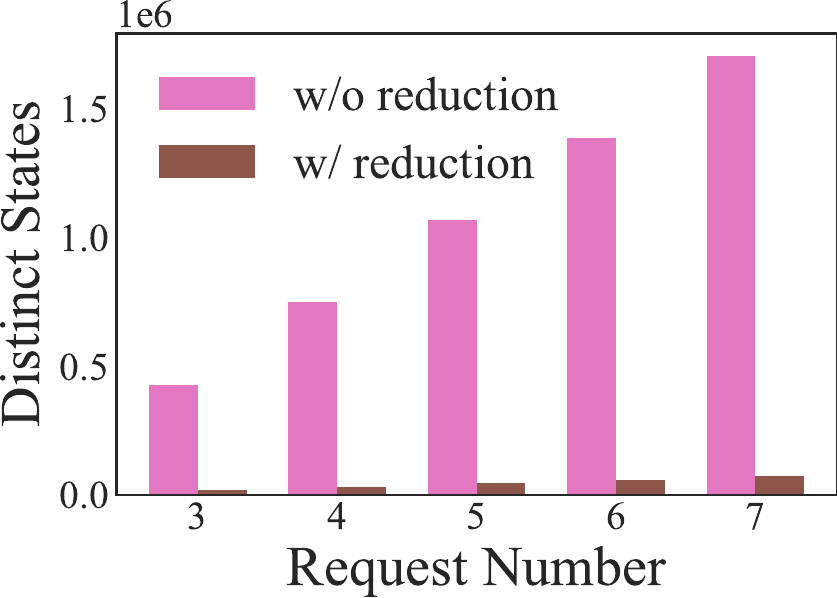}
        }

        \caption{The effect of symmetry reduction.}
        \label{fig:symmetry}
    \end{minipage}
    \begin{minipage}[b]{0.59\textwidth}

        \centering

        \subfloat[Loss.]
        {
        \label{fig:network-nondeterminism-1}
        \includegraphics[width=0.29\linewidth]{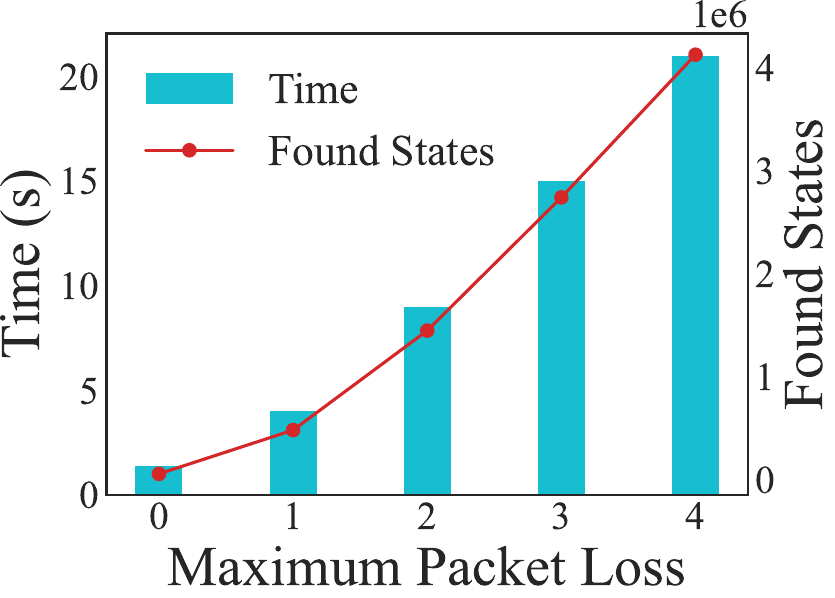}
        }
        \subfloat[Out-of-order.]
        {
        \label{fig:network-nondeterminism-2}
        \includegraphics[width=0.29\linewidth]{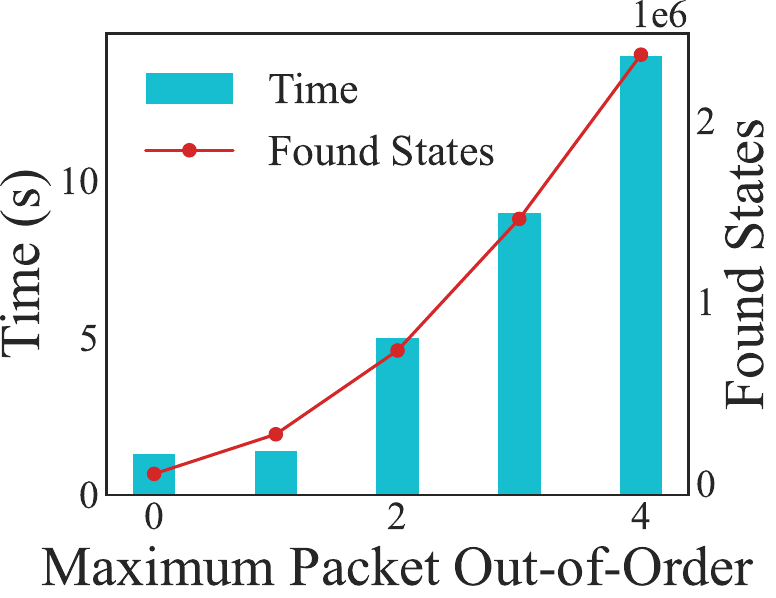}
        }
        \subfloat[Duplication.]
        {
        \label{fig:network-nondeterminism-3}
        \includegraphics[width=0.29\linewidth]{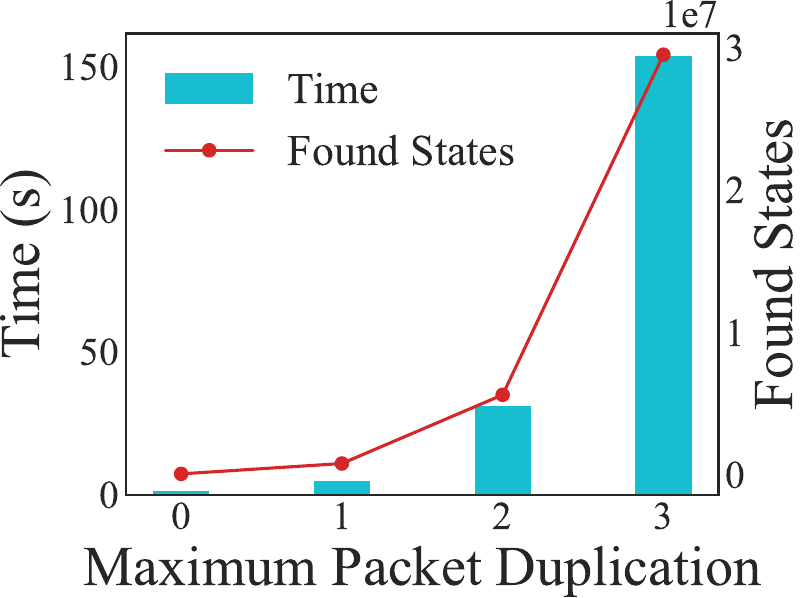}
        }

        \caption{The impact of network nondeterminism.}
        \label{fig:network-nondeterminism}
    \end{minipage}
\end{figure*}

\parab{Effect of optimizations.}
Figure~\ref{fig:symmetry} illustrates the effect of symmetry reduction, using SwitchML as an example. The setup includes a reliable network, a single aggregator, and four clients, disabling packet duplication and property checking, exploring the full state space. As shown, symmetry reduction reduces verification time by 88.1\% and distinct states by 95.5\% on average.

Figure~\ref{fig:network-nondeterminism} shows the impact of network nondeterminism, using NetCache as an example. The setup includes a single client and three requests, disabling property checking, exploring the full state space. We start with a reliable network, gradually relaxing the constraints on packet loss, out-of-order, and duplication. As observed, the size of the state space increases with greater amount of network nondeterminism, leading to a longer verification time. This highlights the necessity of applying appropriate restrictions on network nondeterminism.

\begin{figure}[!ht]
\centering
\includegraphics[width=0.9\linewidth]{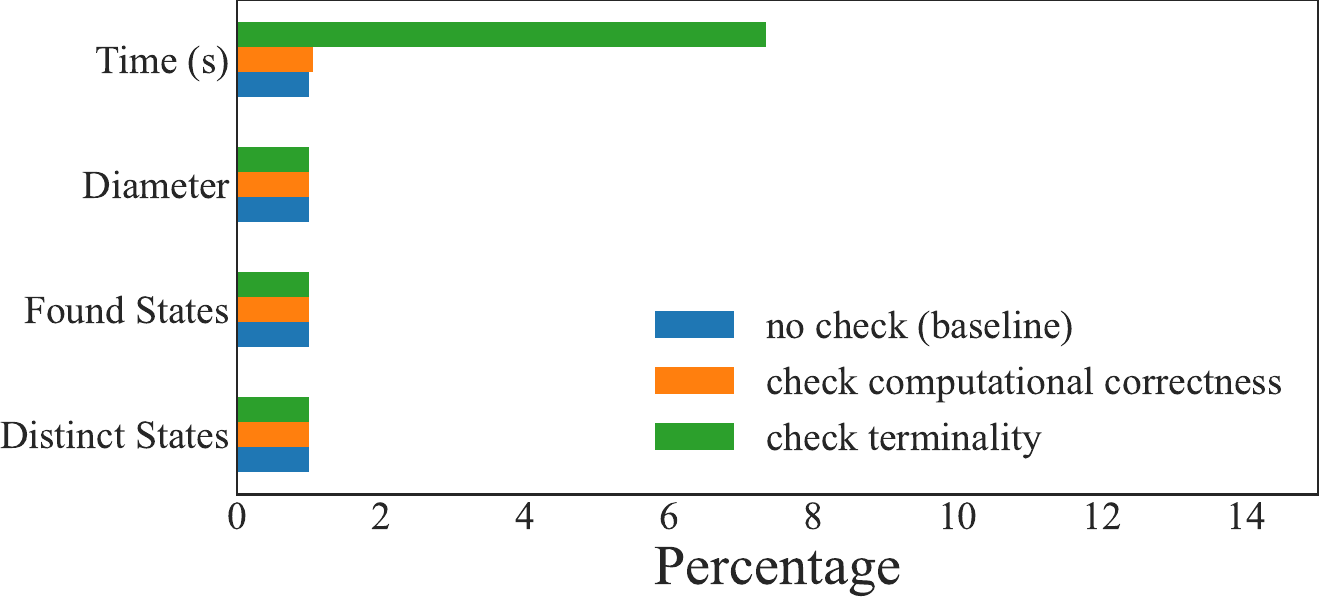}
\caption{The overhead of property checking.}
\label{fig:property}
\end{figure}

\parab{Overhead of property checking.}
Figure~\ref{fig:property} illustrates the overhead of property checking, using SwitchML as an example. The setup includes a reliable network, a single aggregator, two requests, and four clients, disabling packet duplication and symmetry reduction, exploring the full state space. As shown, property checking does not alter the state space itself, but incurs additional verification time. Checking computational correctness adds only a slight overhead, whereas checking terminality significantly increases verification time. This is because the former is a safety property, while the latter is a liveness property~\cite{safety-liveness}. These two types of properties can be transformed into each other under certain conditions. Among all the properties verified in this paper, only terminality is implemented as a liveness property. The definitions of computational correctness and terminality are provided in Section~\ref{ssec:tensor-aggregation-results}.

\section{Verification Result}
\label{sec:use-case}

In total, we modeled \numsystems INC systems with \self. Seven of them exhibit property violations, summarized in Table~\ref{tab:anomalies}; SwitchGNN~\cite{SwitchGNN} passes all checked properties; and EPIC~\cite{EPIC} is discussed in Section~\ref{ssec:epic}.

Each model is constructed from the protocol logic described in the corresponding paper. To validate that our models faithfully capture the relevant protocol behaviors, we reproduce the reported violations in six real implementations and obtain confirmation from the authors of two of the systems.

This section describes the property violations identified, lessons learned, and how \self guided new protocol design. Possible fixes are discussed in Appendix~\ref{sec:fixes}.
We have modeled the proposed fixes and confirmed that the identified violations no longer occur.

\begin{table*}[!ht]
\caption{Property violations in the studied INC systems.}
\label{tab:anomalies}

\resizebox{\linewidth}{!}{
\begin{tabular}{c c c p{0.5\textwidth}}

\toprule
\textbf{System} & \textbf{Network Environment} & \textbf{Property Violation} & \textbf{Explanation} \\
\midrule
\multicolumn{4}{l}{\textbf{Tensor Aggregation}} \\
SwitchML & out-of-order & computational correctness & Out-of-order breaks the implicit assumption that completed requests will not appear in the network \\
ATP & out-of-order & memory leak freedom & A request reaches the switch after aggregation completion, permanently occupying an aggregator  \\
NetReduce & lossy & terminality & CLT update prevents the aggregation of requests in previous messages from completing \\
\midrule
\multicolumn{4}{l}{\textbf{Key-Value Cache}} \\
NetCache & lossy & cache consistency & Particular packet loss and event interleaving expose the risk of replying to the client and updating the switch simultaneously \\
NetCache & reliable & terminality & Cache eviction during a write operation leads to a state inconsistency in the system \\
FarReach & reliable & cache consistency & Particular event interleaving breaks the implicit assumption that read replies from the server carry the latest values \\
\midrule
\multicolumn{4}{l}{\textbf{Lock Management}} \\
NetLock & delayed & lock exclusion & The switch dequeues extra elements upon receiving retransmitted release requests \\
\fisslock & delayed & lock exclusion & Delayed acquisition triggers release and re-acquisition, leading to inconsistency in lock states \\
\bottomrule

\end{tabular}
}
\end{table*}

The INC systems are designed for high-performance datacenters. Yet, few of them explicitly state their network assumptions; the network conditions on which they rely are only implicit in their designs. Our verification results show that most systems operate correctly under reliable networks. However, under unreliable networks, design risks may surface. Identifying such risks during the design phase allows developers to be aware of potential issues, and either fix them before deployment or ensure that the systems are deployed under reliable network environments.

Although modern datacenters employ various techniques to mitigate network unreliability---redundant links, congestion control, and lossless transport such as PFC---packet loss and out-of-order delivery cannot be entirely eliminated. Random bit corruptions, link failures and buffer overflow under extreme congestion still cause packet loss~\cite{DC-Failures}; ECMP-induced path changes and load balancing in RDMA networks still induce packet out-of-order~\cite{IRN, Eunomia}. It is therefore necessary to study system behavior under these unreliable network environments.

\subsection{Tensor Aggregation Systems}
\label{ssec:tensor-aggregation-results}

\parab{System model.}
We model a tensor aggregation system using a star topology, where a switch connects multiple clients and a possible parameter server (PS). Each client has $N$ tensors to be aggregated. Clients send each tensor in a request, expecting to receive the sum of the corresponding tensors from all clients, i.e., performing an all-reduce operation. INC protocols allow the switch to intercept requests, complete the aggregation within the switch, and broadcast the results back to the clients, optionally falling back to the PS under certain conditions. We choose SwitchML~\cite{SwitchML}, ATP~\cite{ATP}, and NetReduce~\cite{NetReduce} as representative protocols.

\parab{Correctness properties.} We specify the following correctness properties with \sysname.

$\bullet$ \textit{Terminality}, which requires the system to eventually reach a termination state (i.e., no deadlock or livelock). Terminality is expressed as the LTL formula ${<>}({active\_nodes}=\emptyset)$, where $<>$ is the LTL operator ``eventually''.

$\bullet$ \textit{Computational correctness}, which requires the aggregation result received by clients to be the exact sum of corresponding aggregated values. We precompute the correct aggregation results and check computational correctness when clients receive aggregation results.

$\bullet$ \textit{Memory leak freedom} for ATP only, which requires a task to occupy no aggregator when it terminates. As we only focus on a specific task, memory leak freedom is checked on system termination, expressed as ${active\_nodes}=\emptyset\to\forall i({aggregators}[i].{id}\ne {TASK\_ID})$.

Table~\ref{tab:anomalies} summarizes the identified violations in the three systems. Protocol-specific models and violation traces are presented in Appendix~\ref{sec:tensor-aggregation-details} due to space limitations.

\subsection{Key-Value Cache Systems}
\label{ssec:key-value-results}

\parab{System model.} We model a key-value cache system using a star topology, where a switch connects multiple clients and a storage server; the system also incorporates a controller connected to the switch and the server. As discussed in Section~\ref{sec:optimization:item}, we only focus on a specific item ( key-value pair). Clients send requests to the storage server to read or write to the item. INC protocols allow the switch to admit an item into the cache, serve read/write requests, and evict an item from the cache under certain conditions. We choose NetCache~\cite{NetCache} and FarReach~\cite{FarReach} as representative protocols.

NetCache implements a write-through cache on the switch. When the switch receives a request on an uncached or invalidated item, it forwards the request to the server. Otherwise, if it is a read, the switch directly replies to the client; if it is a write, the switch invalidates the item and forwards it to the server. Upon receiving the write on a cached item, the server updates its database, then replies to the client and updates the switch cache simultaneously. To ensure cache consistency, the server blocks subsequent writes until it confirms that the switch cache has been updated.
The controller helps the switch decide which cache items to evict and where to admit new ones.
During cache eviction, the switch simply deletes the item since it is a write-through cache.

FarReach implements a write-back cache on the switch. Unlike NetCache, when the switch receives a write on a valid cached item, it directly updates the cache and replies to the client. To achieve non-blocking cache admission/eviction while ensuring cache consistency, FarReach introduces a complex protocol. To admit a new item, the switch first marks it as ``outdated'' and treats it as invalid. When receiving a write request or a read reply (from the server) on the item, the switch extracts the latest value of the item from the packet and updates the cache. The item is then marked as ``latest'' and used for handling requests, ending the admission. %

\parab{Correctness properties.} We specify the following correctness properties with \sysname.

$\bullet$ \textit{Terminality}, the same as described in Section~\ref{ssec:tensor-aggregation-results}.

$\bullet$ \textit{Cache consistency}. Since in-network caching serves as a transparent layer, its primary role is to maintain the consistency provided by the underlying key-value store. Informally, an item is consistent if every read not concurrent with any write returns one of the most recent written values~\cite{kv-consistency}.
The formal definition, an illustrative example, and the checking algorithm are provided in Appendix~\ref{sec:cache-consistency}.

\begin{figure}[!ht]
\centering
\includegraphics[width=0.38\textwidth]{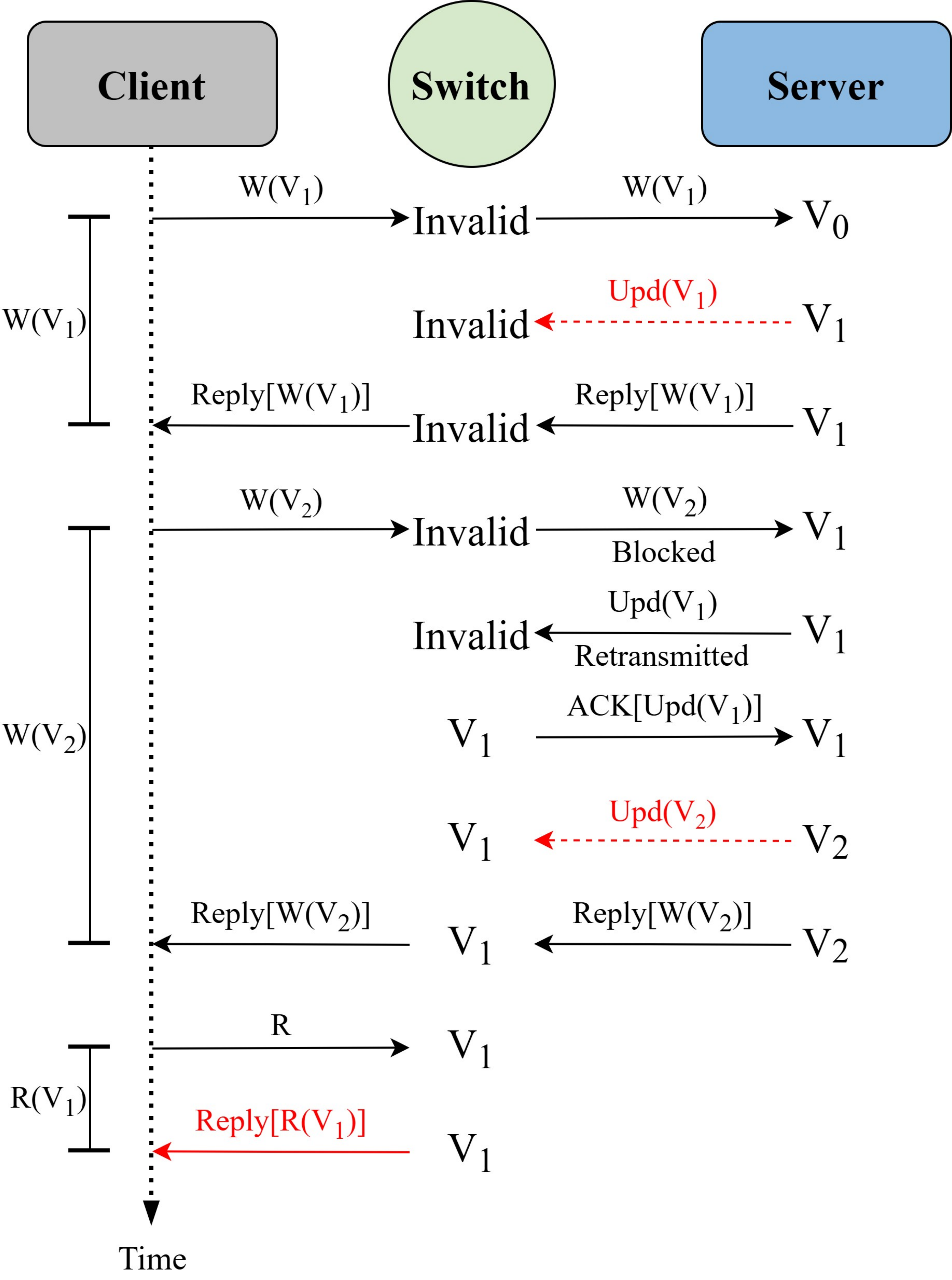}
\caption{An execution trace of cache consistency violation in NetCache. All operations perform on the same item, which is initially $V_0$ and cached in the switch. Dashed lines represent packet loss. Important events are marked in red.}
\label{fig:netcache-coherence-trace}
\end{figure}

\parab{Property violations.}

$\bullet$ \textit{Cache consistency violation in NetCache.} Figure~\ref{fig:netcache-coherence-trace} illustrates an execution trace of cache consistency violation in NetCache. The client initiates a request to write the item to $V_1$. The switch, recognizing the item is cached, invalidates it and forwards the request to the server. The server then updates its local copy of the item to $V_1$, replies to the client, and simultaneously updates the switch cache. However, the update packet is lost. Upon receiving the reply, the client sends a subsequent request to write the item to $V_2$. Since the item is invalid, the switch forwards it to the server. The server blocks it, as the previous update has not yet completed. After a timeout, the server retransmits the previous update. The switch processes it by updating the cache, validating the item, and acknowledging the server. The server proceeds with the blocked write, modifying the local copy, replying to the client, and updating the switch cache. Unfortunately, the update packet is lost again. Upon receiving the reply, the client sends a read and is replied to by the switch with $V_1$. This violates cache consistency since the read is not concurrent with any write but does not get the latest value $V_2$.

$\bullet$ \textit{Terminality violation in NetCache.} NetCache does not provide a detailed discussion on the simultaneous occurrence of multiple events. Without special handling, the following trace is possible. The switch receives a write request, invalidates the corresponding item, and forwards it to the server. At this point, the switch receives a notification from the controller to evict the item, so the switch deletes it. Consequently, when the switch receives the update packet from the server, it discards the packet because there is no such item in the cache. This results in the write request never being completed, and all subsequent writes are indefinitely blocked.

\begin{figure}[!ht]
\centering
\includegraphics[width=0.44\textwidth]{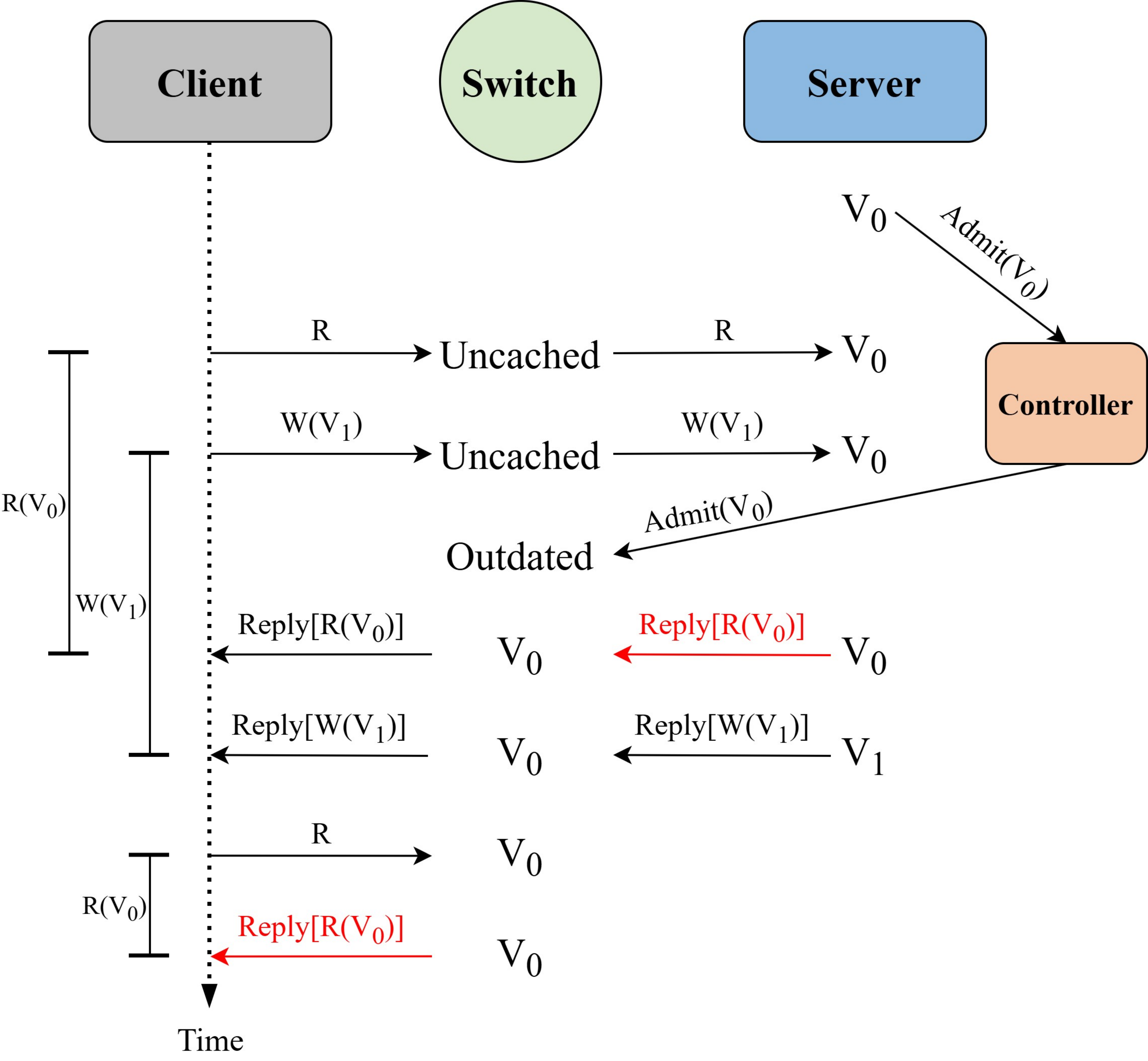}
\caption{An execution trace of cache consistency violation in FarReach. All operations perform on the same item, which is initially $V_0$ and uncached. Important events are marked in red.}
\label{fig:farreach-coherence-trace}
\end{figure}

$\bullet$ \textit{Cache consistency violation in FarReach.} Figure~\ref{fig:farreach-coherence-trace} illustrates an execution trace of cache consistency violation in FarReach concerning admission. The item is initially uncached. When an admission is triggered, the server sends the latest value $V_0$ to the controller. The controller then notifies the switch to admit the item. Before arrival, the switch processes a read and a write request, forwarding them to the server. Upon receiving the admission notification, the switch inserts $V_0$ into the cache, marking it as outdated. Later, the switch receives the reply to the read. Believing it carries the latest value, the switch updates the cached item to $V_0$ and marks it as the latest. Following the replies to both the read and write requests, the client initiates another read and is replied to by the switch with $V_0$. This violates cache consistency since the read is not concurrent with any write but does not get the latest value $V_1$.

\subsection{Lock Management Systems}

We modeled NetLock~\cite{NetLock} and \fisslock~\cite{FISSLOCK}, and identified their violations of lock exclusion under delayed networks, where packets are delivered reliably but may experience delays that trigger timeouts and retransmissions. The violation in NetLock is conceptually straightforward to fix, but is challenging to implement using programmable switches. It is difficult to provide a simple fix to the violations in \fisslock. Details are presented in Appendix~\ref{sec:lock-management-results} (the supplementary material) due to space limitations.

\subsection{Reproduction in Real System}

To demonstrate that the design risks identified by \self indeed exist in real implementations, we conduct reproduction experiments using the open-source code of the aforementioned systems.

The main challenge lies in that the identified violations are triggered only under specific conditions, making them difficult to reproduce directly in a physical environment. To address this, we deploy each node of the system (including clients, programmable switches, and servers) inside individual Docker containers. We further introduce an additional proxy node responsible for forwarding network traffic among nodes. The proxy connects to all other nodes via virtual Ethernet pairs. This setup preserves the original system behavior while allowing us to manipulate the network through the proxy node.

Following the violation traces reported by \self, we inject packet loss and delays (consequently packet out-of-order) into the network via the proxy node. By analyzing the system logs, we confirm that all six systems (excluding NetReduce, which is close-sourced) indeed exhibit the violations we identified.
We successfully contacted the authors of ATP and NetReduce, who acknowledged that the risks we identified were caused by oversights during the design phase.
We have open-sourced the reproduction code for the six systems on GitHub, including scripts for one-click container build and experiment execution.

\subsection{Guiding New Protocol Design}
\label{ssec:epic}

\self has been adopted by EPIC~\cite{EPIC}, an INC collective communication protocol published at SIGCOMM 2026, to guide its protocol design. \self was used to model the polymorphic data plane modes proposed by EPIC and verify computational correctness and protocol terminality under various network environments. \self revealed a critical pitfall in buffer management, which was fixed during the design process. This experience demonstrates that \self is effective not only for uncovering risks in existing protocols but also for guiding the development of new ones.

\subsection{Lessons Learned}

From the violations across these INC systems, we distill three recurring patterns.

\parab{Stale packets polluting switch-side state.}
Violations in several systems (SwitchML, ATP, NetReduce, NetLock, and FissLock) occur because the switch processes stale packets (e.g., packets from previous rounds or retransmissions) incorrectly, corrupting its stateful entries. Designers should \textit{ensure that the switch can identify stale packets and handle them properly}, for example, by tracking packet sequence numbers.

\parab{Concurrent events racing without coordination.}
Violations in several systems (NetCache and FarReach) arise from uncoordinated concurrent events (e.g., eviction racing with an ongoing write, or admission racing with in-flight replies). Designers should \textit{explicitly handle the interplay between concurrent protocol events}, rather than implicitly assuming each event occurs in isolation.

\parab{Network unreliability amplifying design risks.}
Design risks may remain latent under ideal conditions, but network delays, packet loss, or out-of-order delivery can turn them into real violations. Most violations we found require only a single such event to trigger. Designers should \textit{systematically verify protocol correctness under diverse network conditions}, not just under ideal delivery assumptions.

\section{Related Work}
\label{sec:related-work}

\parab{In-network computing.} High-speed programmable swi\-tches have enabled various INC techniques to offload computation from end hosts to the network. ATP~\cite{ATP} uses top-of-rack switches to aggregate tensors and hugely accelerates distributed training throughput. NetCache~\cite{NetCache} implements an on-path key-value cache in switches to balance the load across storage nodes. \fisslock~\cite{FISSLOCK} decouples lock management into grant decision and participant maintenance, supporting over one million locks on one switch. %

These works focus mainly on improving system performance and somewhat neglect the correctness under unreliable networks. %
Only few protocols~\cite{NetChain,RedPlane} have seriously considered correctness issues and verified themselves using formal methods. We believe that INC applications should pay more attention to correctness, and provide \self as an easy-to-use, general-purpose verification tool.

\parab{INC verification.} Programmable switches have brought new topics to network verification. ASSERT-P4~\cite{ASSERT-P4} verifies P4 programs annotated with assertions based on symbolic execution. Aquila~\cite{Aquila} presents a verification system for production-scale programmable data planes.
There are also testing tools applicable to INC systems,
such as P6~\cite{P6} and Chimera~\cite{Chimera}.
They differ from \self in two respects. First, \self finds design-level risks
(e.g., cache inconsistency) in high-level specifications, whereas
these tools find implementation-level bugs (e.g., out-of-bounds
array accesses) in executable programs. Second, successful
verification proves the specified properties, whereas testing cannot establish the absence of bugs.

\parab{Model checking in network verification.} Model checking has been widely used in network verification.  Musuvath \textit{et al.} \cite{MC-TCP} propose model checking techniques and finds four errors in Linux TCP/IP implementation. NetSMC~\cite{NetSMC} adopts a one-packet-at-a-time network model and verifies routing policies in stateful networks with symbolic model checking. Plankton~\cite{Plankton} combines equivalence partitioning with model checking to perform efficient network configuration verification. We differ from these works in that we target design correctness concerning properties required by applications.

\section{Conclusion}
\label{sec:conclusion}

We present \self, the first general-purpose tool for verifying INC systems. \self provides a high-level specification language and configurable network environments to help developers better understand the behavior of the system. We modeled INC systems from various application domains and identified their design risks with \self in seconds.
We believe that formal modeling and verification are an essential step in driving the deployment of INC systems.

\section*{Acknowledgment}
We would like to thank the anonymous reviewers and the shepherd for their insightful comments and suggestions.
Wenfei Wu is supported by ``The Fundamental Research Funds for the Central Universities, Peking University''.

\clearpage \newpage
\bibliographystyle{IEEEtran}
\bibliography{reference}

\begin{thebibliography}{10}
\providecommand{\url}[1]{#1}
\csname url@samestyle\endcsname
\providecommand{\newblock}{\relax}
\providecommand{\bibinfo}[2]{#2}
\providecommand{\BIBentrySTDinterwordspacing}{\spaceskip=0pt\relax}
\providecommand{\BIBentryALTinterwordstretchfactor}{4}
\providecommand{\BIBentryALTinterwordspacing}{\spaceskip=\fontdimen2\font plus
\BIBentryALTinterwordstretchfactor\fontdimen3\font minus
  \fontdimen4\font\relax}
\providecommand{\BIBforeignlanguage}[2]{{%
\expandafter\ifx\csname l@#1\endcsname\relax
\typeout{** WARNING: IEEEtran.bst: No hyphenation pattern has been}%
\typeout{** loaded for the language `#1'. Using the pattern for}%
\typeout{** the default language instead.}%
\else
\language=\csname l@#1\endcsname
\fi
#2}}
\providecommand{\BIBdecl}{\relax}
\BIBdecl

\bibitem{Tofino}
Intel, ``{Barefoot} {Tofino},''
  \url{https://www.intel.com/content/www/us/en/products/network-io/programmable-ethernet-switch.html\#tofino},
  2020.

\bibitem{Cisco}
Cisco, ``{Cisco Silicon One},''
  \url{https://blogs.cisco.com/sp/one-silicon-one-experience-multiple-roles},
  2019.

\bibitem{Trident}
\BIBentryALTinterwordspacing
Broadcom, ``Broadcom trident,'' 2025. [Online]. Available:
  \url{https://www.broadcom.com/products/ethernet-connectivity/switching/strataxgs/smarttor}
\BIBentrySTDinterwordspacing

\bibitem{Trio}
M.~Yang, A.~Baban, V.~Kugel, J.~Libby, S.~Mackie, S.~S.~R. Kananda, C.-H. Wu,
  and M.~Ghobadi, ``Using trio: Juniper networks' programmable chipset-for
  emerging in-network applications,'' in \emph{Proceedings of ACM SIGCOMM},
  2022, pp. 633--648.

\bibitem{NetEngine}
\BIBentryALTinterwordspacing
Huawei, ``Huawei netengine,'' 2025. [Online]. Available:
  \url{https://e.huawei.com/en/products/routers/ne8000}
\BIBentrySTDinterwordspacing

\bibitem{SwitchML}
\BIBentryALTinterwordspacing
A.~Sapio, M.~Canini, C.-Y. Ho, J.~Nelson, P.~Kalnis, C.~Kim, A.~Krishnamurthy,
  M.~Moshref, D.~Ports, and P.~Richtarik, ``Scaling distributed machine
  learning with in-network aggregation,'' in \emph{18th USENIX Symposium on
  Networked Systems Design and Implementation (NSDI 21)}.\hskip 1em plus 0.5em
  minus 0.4em\relax USENIX Association, Apr. 2021, pp. 785--808. [Online].
  Available: \url{https://www.usenix.org/conference/nsdi21/presentation/sapio}
\BIBentrySTDinterwordspacing

\bibitem{ATP}
\BIBentryALTinterwordspacing
C.~Lao, Y.~Le, K.~Mahajan, Y.~Chen, W.~Wu, A.~Akella, and M.~Swift, ``{ATP}:
  In-network aggregation for multi-tenant learning,'' in \emph{18th USENIX
  Symposium on Networked Systems Design and Implementation (NSDI 21)}.\hskip
  1em plus 0.5em minus 0.4em\relax USENIX Association, Apr. 2021, pp. 741--761.
  [Online]. Available:
  \url{https://www.usenix.org/conference/nsdi21/presentation/lao}
\BIBentrySTDinterwordspacing

\bibitem{DSA}
H.~Wang, Y.~Qin, C.~Lao, Y.~Le, W.~Wu, and K.~Chen, ``Preemptive switch memory
  usage to accelerate training jobs with shared in-network aggregation,'' in
  \emph{2023 IEEE 31st International Conference on Network Protocols (ICNP)},
  2023, pp. 1--12.

\bibitem{NetReduce}
\BIBentryALTinterwordspacing
S.~Liu, Q.~Wang, J.~Zhang, W.~Wu, Q.~Lin, Y.~Liu, M.~Xu, M.~Canini, R.~C.~C.
  Cheung, and J.~He, ``In-network aggregation with transport transparency for
  distributed training,'' in \emph{Proceedings of the 28th ACM International
  Conference on Architectural Support for Programming Languages and Operating
  Systems, Volume 3}, ser. ASPLOS 2023.\hskip 1em plus 0.5em minus 0.4em\relax
  New York, NY, USA: Association for Computing Machinery, 2023, pp. 376--391.
  [Online]. Available: \url{https://doi.org/10.1145/3582016.3582037}
\BIBentrySTDinterwordspacing

\bibitem{NetCache}
\BIBentryALTinterwordspacing
X.~Jin, X.~Li, H.~Zhang, R.~Soul\'{e}, J.~Lee, N.~Foster, C.~Kim, and
  I.~Stoica, ``{NetCache}: Balancing key-value stores with fast in-network
  caching,'' in \emph{Proceedings of the 26th Symposium on Operating Systems
  Principles}, ser. SOSP '17.\hskip 1em plus 0.5em minus 0.4em\relax New York,
  NY, USA: Association for Computing Machinery, 2017, pp. 121--136. [Online].
  Available: \url{https://doi.org/10.1145/3132747.3132764}
\BIBentrySTDinterwordspacing

\bibitem{DistCache}
\BIBentryALTinterwordspacing
Z.~Liu, Z.~Bai, Z.~Liu, X.~Li, C.~Kim, V.~Braverman, X.~Jin, and I.~Stoica,
  ``{DistCache}: Provable load balancing for {large-scale} storage systems with
  distributed caching,'' in \emph{17th USENIX Conference on File and Storage
  Technologies (FAST 19)}.\hskip 1em plus 0.5em minus 0.4em\relax Boston, MA:
  USENIX Association, Feb. 2019, pp. 143--157. [Online]. Available:
  \url{https://www.usenix.org/conference/fast19/presentation/liu}
\BIBentrySTDinterwordspacing

\bibitem{FarReach}
\BIBentryALTinterwordspacing
S.~Sheng, H.~Puyang, Q.~Huang, L.~Tang, and P.~P.~C. Lee, ``{FarReach}:
  Write-back caching in programmable switches,'' in \emph{2023 USENIX Annual
  Technical Conference (USENIX ATC 23)}.\hskip 1em plus 0.5em minus 0.4em\relax
  Boston, MA: USENIX Association, Jul. 2023, pp. 571--584. [Online]. Available:
  \url{https://www.usenix.org/conference/atc23/presentation/sheng}
\BIBentrySTDinterwordspacing

\bibitem{Seer}
\BIBentryALTinterwordspacing
J.~Lei and V.~Shrivastav, ``{Seer}: Enabling {future-aware} online caching in
  networked systems,'' in \emph{21st USENIX Symposium on Networked Systems
  Design and Implementation (NSDI 24)}.\hskip 1em plus 0.5em minus 0.4em\relax
  Santa Clara, CA: USENIX Association, Apr. 2024, pp. 635--649. [Online].
  Available: \url{https://www.usenix.org/conference/nsdi24/presentation/lei}
\BIBentrySTDinterwordspacing

\bibitem{NetLock}
\BIBentryALTinterwordspacing
Z.~Yu, Y.~Zhang, V.~Braverman, M.~Chowdhury, and X.~Jin, ``{NetLock}: Fast,
  centralized lock management using programmable switches,'' in
  \emph{Proceedings of the Annual Conference of the ACM Special Interest Group
  on Data Communication on the Applications, Technologies, Architectures, and
  Protocols for Computer Communication}, ser. SIGCOMM '20.\hskip 1em plus 0.5em
  minus 0.4em\relax New York, NY, USA: Association for Computing Machinery,
  2020, pp. 126--138. [Online]. Available:
  \url{https://doi.org/10.1145/3387514.3405857}
\BIBentrySTDinterwordspacing

\bibitem{FISSLOCK}
\BIBentryALTinterwordspacing
H.~Zhang, K.~Cheng, R.~Chen, and H.~Chen, ``Fast and scalable in-network lock
  management using lock fission,'' in \emph{18th USENIX Symposium on Operating
  Systems Design and Implementation (OSDI 24)}.\hskip 1em plus 0.5em minus
  0.4em\relax Santa Clara, CA: USENIX Association, Jul. 2024, pp. 251--268.
  [Online]. Available:
  \url{https://www.usenix.org/conference/osdi24/presentation/zhang-hanze}
\BIBentrySTDinterwordspacing

\bibitem{SwitchV2P}
\BIBentryALTinterwordspacing
L.~Zeno, A.~Chen, and M.~Silberstein, ``In-network address caching for virtual
  networks,'' in \emph{Proceedings of the ACM SIGCOMM 2024 Conference}, ser.
  ACM SIGCOMM '24.\hskip 1em plus 0.5em minus 0.4em\relax New York, NY, USA:
  Association for Computing Machinery, 2024, pp. 735--749. [Online]. Available:
  \url{https://doi.org/10.1145/3651890.3672213}
\BIBentrySTDinterwordspacing

\bibitem{RackSched}
\BIBentryALTinterwordspacing
H.~Zhu, K.~Kaffes, Z.~Chen, Z.~Liu, C.~Kozyrakis, I.~Stoica, and X.~Jin,
  ``{RackSched}: A {microsecond-scale} scheduler for {rack-scale} computers,''
  in \emph{14th USENIX Symposium on Operating Systems Design and Implementation
  (OSDI 20)}.\hskip 1em plus 0.5em minus 0.4em\relax USENIX Association, Nov.
  2020, pp. 1225--1240. [Online]. Available:
  \url{https://www.usenix.org/conference/osdi20/presentation/zhu}
\BIBentrySTDinterwordspacing

\bibitem{Horus}
\BIBentryALTinterwordspacing
P.~Yassini, K.~Diab, S.~Mahloujifar, and M.~Hefeeda, ``{Horus}: Granular
  {in-network} task scheduler for cloud datacenters,'' in \emph{21st USENIX
  Symposium on Networked Systems Design and Implementation (NSDI 24)}.\hskip
  1em plus 0.5em minus 0.4em\relax Santa Clara, CA: USENIX Association, Apr.
  2024, pp. 1--22. [Online]. Available:
  \url{https://www.usenix.org/conference/nsdi24/presentation/yassini}
\BIBentrySTDinterwordspacing

\bibitem{P4xos}
H.~T. Dang, P.~Bressana, H.~Wang, K.~S. Lee, N.~Zilberman, H.~Weatherspoon,
  M.~Canini, F.~Pedone, and R.~Soulé, ``{P4xos}: Consensus as a network
  service,'' \emph{IEEE/ACM Transactions on Networking}, vol.~28, no.~4, pp.
  1726--1738, 2020.

\bibitem{NetPaxos}
\BIBentryALTinterwordspacing
H.~T. Dang, D.~Sciascia, M.~Canini, F.~Pedone, and R.~Soul\'{e}, ``{NetPaxos}:
  Consensus at network speed,'' in \emph{Proceedings of the 1st ACM SIGCOMM
  Symposium on Software Defined Networking Research}, ser. SOSR '15.\hskip 1em
  plus 0.5em minus 0.4em\relax New York, NY, USA: Association for Computing
  Machinery, 2015. [Online]. Available:
  \url{https://doi.org/10.1145/2774993.2774999}
\BIBentrySTDinterwordspacing

\bibitem{NOPaxos}
\BIBentryALTinterwordspacing
J.~Li, E.~Michael, N.~K. Sharma, A.~Szekeres, and D.~R.~K. Ports, ``Just say
  {NO} to {Paxos} overhead: Replacing consensus with network ordering,'' in
  \emph{12th USENIX Symposium on Operating Systems Design and Implementation
  (OSDI 16)}.\hskip 1em plus 0.5em minus 0.4em\relax Savannah, GA: USENIX
  Association, Nov. 2016, pp. 467--483. [Online]. Available:
  \url{https://www.usenix.org/conference/osdi16/technical-sessions/presentation/li}
\BIBentrySTDinterwordspacing

\bibitem{P4-Survey}
E.~F. Kfoury, J.~Crichigno, and E.~Bou-Harb, ``An exhaustive survey on {P4}
  programmable data plane switches: Taxonomy, applications, challenges, and
  future trends,'' \emph{IEEE Access}, vol.~9, pp. 87\,094--87\,155, 2021.

\bibitem{RedPlane}
\BIBentryALTinterwordspacing
D.~Kim, J.~Nelson, D.~R.~K. Ports, V.~Sekar, and S.~Seshan, ``Redplane:
  Enabling fault-tolerant stateful in-switch applications,'' in
  \emph{Proceedings of the 2021 ACM SIGCOMM 2021 Conference}, ser. SIGCOMM
  '21.\hskip 1em plus 0.5em minus 0.4em\relax New York, NY, USA: Association
  for Computing Machinery, 2021, pp. 223--244. [Online]. Available:
  \url{https://doi.org/10.1145/3452296.3472905}
\BIBentrySTDinterwordspacing

\bibitem{YCSB}
\BIBentryALTinterwordspacing
B.~F. Cooper, A.~Silberstein, E.~Tam, R.~Ramakrishnan, and R.~Sears,
  ``Benchmarking cloud serving systems with {YCSB},'' in \emph{Proceedings of
  the 1st ACM Symposium on Cloud Computing}, ser. SoCC '10.\hskip 1em plus
  0.5em minus 0.4em\relax New York, NY, USA: Association for Computing
  Machinery, 2010, pp. 143--154. [Online]. Available:
  \url{https://doi.org/10.1145/1807128.1807152}
\BIBentrySTDinterwordspacing

\bibitem{ns3}
ns~3~Consortium, ``ns-3: Network simulator 3,'' \url{https://www.nsnam.org/},
  2006.

\bibitem{Reachability1}
G.~Xie, J.~Zhan, D.~Maltz, H.~Zhang, A.~Greenberg, G.~Hjalmtysson, and
  J.~Rexford, ``On static reachability analysis of {IP} networks,'' in
  \emph{Proceedings IEEE 24th Annual Joint Conference of the IEEE Computer and
  Communications Societies.}, vol.~3, 2005, pp. 2170--2183 vol. 3.

\bibitem{Reachability2}
A.~R. Khakpour and A.~X. Liu, ``Quantifying and querying network
  reachability,'' in \emph{2010 IEEE 30th International Conference on
  Distributed Computing Systems}, 2010, pp. 817--826.

\bibitem{HSA}
P.~Kazemian, G.~Varghese, and N.~McKeown, ``Header space analysis: Static
  checking for networks,'' in \emph{Proceedings of the 9th USENIX Conference on
  Networked Systems Design and Implementation}, ser. NSDI'12.\hskip 1em plus
  0.5em minus 0.4em\relax USA: USENIX Association, 2012, p.~9.

\bibitem{VeriFlow}
\BIBentryALTinterwordspacing
A.~Khurshid, X.~Zou, W.~Zhou, M.~Caesar, and P.~B. Godfrey, ``{Veriflow}:
  Verifying {network-wide} invariants in real time,'' in \emph{10th USENIX
  Symposium on Networked Systems Design and Implementation (NSDI 13)}.\hskip
  1em plus 0.5em minus 0.4em\relax Lombard, IL: USENIX Association, Apr. 2013,
  pp. 15--27. [Online]. Available:
  \url{https://www.usenix.org/conference/nsdi13/technical-sessions/presentation/khurshid}
\BIBentrySTDinterwordspacing

\bibitem{Isolation1}
\BIBentryALTinterwordspacing
S.~Gutz, A.~Story, C.~Schlesinger, and N.~Foster, ``Splendid isolation: A slice
  abstraction for software-defined networks,'' in \emph{Proceedings of the
  First Workshop on Hot Topics in Software Defined Networks}, ser. HotSDN
  '12.\hskip 1em plus 0.5em minus 0.4em\relax New York, NY, USA: Association
  for Computing Machinery, 2012, pp. 79--84. [Online]. Available:
  \url{https://doi.org/10.1145/2342441.2342458}
\BIBentrySTDinterwordspacing

\bibitem{AP-Verifier}
H.~Yang and S.~S. Lam, ``Real-time verification of network properties using
  atomic predicates,'' \emph{IEEE/ACM Transactions on Networking}, vol.~24,
  no.~2, pp. 887--900, 2016.

\bibitem{NetSMC}
\BIBentryALTinterwordspacing
Y.~Yuan, S.-J. Moon, S.~Uppal, L.~Jia, and V.~Sekar, ``{NetSMC}: A custom
  symbolic model checker for stateful network verification,'' in \emph{17th
  USENIX Symposium on Networked Systems Design and Implementation (NSDI
  20)}.\hskip 1em plus 0.5em minus 0.4em\relax Santa Clara, CA: USENIX
  Association, Feb. 2020, pp. 181--200. [Online]. Available:
  \url{https://www.usenix.org/conference/nsdi20/presentation/yuan}
\BIBentrySTDinterwordspacing

\bibitem{DNS-V}
\BIBentryALTinterwordspacing
N.~Zheng, M.~Liu, Y.~Xiang, L.~Song, D.~Li, F.~Han, N.~Wang, Y.~Ma, Z.~Liang,
  D.~Cai, E.~Zhai, X.~Liu, and X.~Jin, ``Automated verification of an
  in-production {DNS} authoritative engine,'' in \emph{Proceedings of the 29th
  Symposium on Operating Systems Principles}, ser. SOSP '23.\hskip 1em plus
  0.5em minus 0.4em\relax New York, NY, USA: Association for Computing
  Machinery, 2023, pp. 80--95. [Online]. Available:
  \url{https://doi.org/10.1145/3600006.3613153}
\BIBentrySTDinterwordspacing

\bibitem{P4}
\BIBentryALTinterwordspacing
P.~Bosshart, D.~Daly, G.~Gibb, M.~Izzard, N.~McKeown, J.~Rexford,
  C.~Schlesinger, D.~Talayco, A.~Vahdat, G.~Varghese, and D.~Walker, ``{P4}:
  Programming protocol-independent packet processors,'' \emph{SIGCOMM Comput.
  Commun. Rev.}, vol.~44, no.~3, pp. 87--95, Jul. 2014. [Online]. Available:
  \url{https://doi.org/10.1145/2656877.2656890}
\BIBentrySTDinterwordspacing

\bibitem{p4v}
\BIBentryALTinterwordspacing
J.~Liu, W.~Hallahan, C.~Schlesinger, M.~Sharif, J.~Lee, R.~Soul\'{e}, H.~Wang,
  C.~Ca\c{s}caval, N.~McKeown, and N.~Foster, ``{p4v}: Practical verification
  for programmable data planes,'' in \emph{Proceedings of the 2018 Conference
  of the ACM Special Interest Group on Data Communication}, ser. SIGCOMM
  '18.\hskip 1em plus 0.5em minus 0.4em\relax New York, NY, USA: Association
  for Computing Machinery, 2018, pp. 490--503. [Online]. Available:
  \url{https://doi.org/10.1145/3230543.3230582}
\BIBentrySTDinterwordspacing

\bibitem{Vera}
\BIBentryALTinterwordspacing
R.~Stoenescu, D.~Dumitrescu, M.~Popovici, L.~Negreanu, and C.~Raiciu,
  ``Debugging p4 programs with {Vera},'' in \emph{Proceedings of the 2018
  Conference of the ACM Special Interest Group on Data Communication}, ser.
  SIGCOMM '18.\hskip 1em plus 0.5em minus 0.4em\relax New York, NY, USA:
  Association for Computing Machinery, 2018, pp. 518--532. [Online]. Available:
  \url{https://doi.org/10.1145/3230543.3230548}
\BIBentrySTDinterwordspacing

\bibitem{ASSERT-P4}
\BIBentryALTinterwordspacing
L.~Freire, M.~Neves, L.~Leal, K.~Levchenko, A.~Schaeffer-Filho, and
  M.~Barcellos, ``Uncovering bugs in {P4} programs with assertion-based
  verification,'' in \emph{Proceedings of the Symposium on SDN Research}, ser.
  SOSR '18.\hskip 1em plus 0.5em minus 0.4em\relax New York, NY, USA:
  Association for Computing Machinery, 2018. [Online]. Available:
  \url{https://doi.org/10.1145/3185467.3185499}
\BIBentrySTDinterwordspacing

\bibitem{Aquila}
\BIBentryALTinterwordspacing
B.~Tian, J.~Gao, M.~Liu, E.~Zhai, Y.~Chen, Y.~Zhou, L.~Dai, F.~Yan, M.~Ma,
  M.~Tang, J.~Lu, X.~Wei, H.~H. Liu, M.~Zhang, C.~Tian, and M.~Yu, ``{Aquila}:
  A practically usable verification system for production-scale programmable
  data planes,'' in \emph{Proceedings of the 2021 ACM SIGCOMM 2021 Conference},
  ser. SIGCOMM '21.\hskip 1em plus 0.5em minus 0.4em\relax New York, NY, USA:
  Association for Computing Machinery, 2021, pp. 17--32. [Online]. Available:
  \url{https://doi.org/10.1145/3452296.3472937}
\BIBentrySTDinterwordspacing

\bibitem{SPIN}
\BIBentryALTinterwordspacing
G.~Holzmann, ``\BIBforeignlanguage{en-US}{The model checker {{SPIN}}},''
  \emph{\BIBforeignlanguage{en-US}{IEEE Transactions on Software Engineering}},
  pp. 279--295, May 1997. [Online]. Available:
  \url{http://dx.doi.org/10.1109/32.588521}
\BIBentrySTDinterwordspacing

\bibitem{tlaplus}
L.~Lamport, \emph{Specifying systems: The {TLA+} language and tools for
  hardware and software engineers}.\hskip 1em plus 0.5em minus 0.4em\relax
  Addison-Wesley, Jun. 2002.

\bibitem{SwitchGNN}
\BIBentryALTinterwordspacing
Z.~Li, J.~Huang, Y.~Li, J.~Liu, J.~Zhang, H.~Li, X.~Zhu, S.~Zhou, J.~Shao,
  X.~Lu, Q.~Su, J.~Wang, C.~W. Tan, Y.~Cui, and K.~Chen, ``Accelerating
  distributed graph learning by using collaborative {In-Network} multicast and
  aggregation,'' in \emph{2025 USENIX Annual Technical Conference (USENIX ATC
  25)}.\hskip 1em plus 0.5em minus 0.4em\relax Boston, MA: USENIX Association,
  Jul. 2025, pp. 233--247. [Online]. Available:
  \url{https://www.usenix.org/conference/atc25/presentation/li-zhaoyi}
\BIBentrySTDinterwordspacing

\bibitem{EPIC}
\BIBentryALTinterwordspacing
Y.~Yuan, J.~Nie, T.~Bai, R.~Zhou, S.~Cao, X.~Fan, Y.~Xu, J.~Chen, C.~Zhao,
  N.~Zhang, S.~Fang, J.~Chen, Y.~Chen, J.~Sun, Z.~Wang, X.~Xu, Y.~Zhang,
  Y.~Liu, X.~Yang, J.~Lin, X.~Hu, Y.~Li, C.~Jiang, L.~Xiao, W.~Zhang, J.~Wang,
  W.~Cheng, Y.~Lan, J.~Dong, B.~Fu, and W.~Wu, ``{EPIC}: Abstraction and
  polymorphism of in-network collectives on ethernet,'' 2026. [Online].
  Available: \url{https://arxiv.org/abs/2605.18683}
\BIBentrySTDinterwordspacing

\bibitem{MC1}
\BIBentryALTinterwordspacing
E.~M. Clarke, E.~A. Emerson, and A.~P. Sistla,
  ``\BIBforeignlanguage{en-US}{Automatic verification of finite-state
  concurrent systems using temporal logic specifications},''
  \emph{\BIBforeignlanguage{en-US}{ACM Transactions on Programming Languages
  and Systems}}, pp. 244--263, Apr. 1986. [Online]. Available:
  \url{http://dx.doi.org/10.1145/5397.5399}
\BIBentrySTDinterwordspacing

\bibitem{Theorem-Proving}
C.-L. Chang and R.~C.-T. Lee, \emph{Symbolic logic and mechanical theorem
  proving}, 1st~ed.\hskip 1em plus 0.5em minus 0.4em\relax USA: Academic Press,
  Inc., 1997.

\bibitem{SMT}
\BIBentryALTinterwordspacing
L.~De~Moura and N.~Bj\o{}rner, ``Satisfiability modulo theories: Introduction
  and applications,'' \emph{Commun. ACM}, vol.~54, no.~9, pp. 69--77, Sep.
  2011. [Online]. Available: \url{https://doi.org/10.1145/1995376.1995394}
\BIBentrySTDinterwordspacing

\bibitem{COMST18}
Y.~Li, X.~Yin, Z.~Wang, J.~Yao, X.~Shi, J.~Wu, H.~Zhang, and Q.~Wang, ``A
  survey on network verification and testing with formal methods: Approaches
  and challenges,'' \emph{IEEE Communications Surveys \& Tutorials}, vol.~21,
  no.~1, pp. 940--969, 2019.

\bibitem{verify-paxos}
S.~Chand, Y.~A. Liu, and S.~D. Stoller, ``Formal verification of multi-paxos
  for distributed consensus,'' in \emph{FM 2016: Formal Methods},
  J.~Fitzgerald, C.~Heitmeyer, S.~Gnesi, and A.~Philippou, Eds.\hskip 1em plus
  0.5em minus 0.4em\relax Cham: Springer International Publishing, 2016, pp.
  119--136.

\bibitem{verify-other}
E.~B{\"o}rger, ``Modeling distributed algorithms by abstract state machines
  compared to {Petri} nets,'' in \emph{Abstract State Machines, Alloy, B, TLA,
  VDM, and Z}, M.~Butler, K.-D. Schewe, A.~Mashkoor, and M.~Biro, Eds.\hskip
  1em plus 0.5em minus 0.4em\relax Cham: Springer International Publishing,
  2016, pp. 3--34.

\bibitem{MC-Principle}
C.~Baier and J.-P. Katoen, \emph{Principles of model checking (representation
  and mind series)}.\hskip 1em plus 0.5em minus 0.4em\relax The MIT Press,
  2008.

\bibitem{MC-Symmetry}
\BIBentryALTinterwordspacing
E.~A. Emerson and A.~P. Sistla, ``Symmetry and model checking,'' \emph{Form.
  Methods Syst. Des.}, vol.~9, no. 1--2, pp. 105--131, Aug. 1996. [Online].
  Available: \url{https://doi.org/10.1007/BF00625970}
\BIBentrySTDinterwordspacing

\bibitem{Coherence-TLA}
\BIBentryALTinterwordspacing
R.~Joshi, L.~Lamport, J.~Matthews, S.~Tasiran, M.~Tuttle, and Y.~Yu, ``Checking
  cache-coherence protocols with tla+,'' \emph{Form. Methods Syst. Des.},
  vol.~22, no.~2, pp. 125--131, Mar. 2003. [Online]. Available:
  \url{https://doi.org/10.1023/A:1022969405325}
\BIBentrySTDinterwordspacing

\bibitem{flex-bison}
J.~Levine, \emph{{Flex} \& {Bison}: Text processing tools}.\hskip 1em plus
  0.5em minus 0.4em\relax " O'Reilly Media, Inc.", 2009.

\bibitem{PlusCal}
L.~Lamport, ``The {PlusCal} algorithm language,'' in \emph{Theoretical aspects
  of computing - ICTAC 2009}, M.~Leucker and C.~Morgan, Eds.\hskip 1em plus
  0.5em minus 0.4em\relax Berlin, Heidelberg: Springer Berlin Heidelberg, 2009,
  pp. 36--60.

\bibitem{safety-liveness}
\BIBentryALTinterwordspacing
Z.~Manna and A.~Pnueli, ``A hierarchy of temporal properties (invited paper,
  1989),'' in \emph{Proceedings of the Ninth Annual ACM Symposium on Principles
  of Distributed Computing}, ser. PODC '90.\hskip 1em plus 0.5em minus
  0.4em\relax New York, NY, USA: Association for Computing Machinery, 1990, pp.
  377--410. [Online]. Available: \url{https://doi.org/10.1145/93385.93442}
\BIBentrySTDinterwordspacing

\bibitem{DC-Failures}
\BIBentryALTinterwordspacing
P.~Gill, N.~Jain, and N.~Nagappan, ``Understanding network failures in data
  centers: measurement, analysis, and implications,'' in \emph{Proceedings of
  the ACM SIGCOMM 2011 Conference}, ser. SIGCOMM '11.\hskip 1em plus 0.5em
  minus 0.4em\relax New York, NY, USA: Association for Computing Machinery,
  2011, pp. 350--361. [Online]. Available:
  \url{https://doi.org/10.1145/2018436.2018477}
\BIBentrySTDinterwordspacing

\bibitem{IRN}
\BIBentryALTinterwordspacing
R.~Mittal, A.~Shpiner, A.~Panda, E.~Zahavi, A.~Krishnamurthy, S.~Ratnasamy, and
  S.~Shenker, ``Revisiting network support for rdma,'' in \emph{Proceedings of
  the 2018 Conference of the ACM Special Interest Group on Data Communication},
  ser. SIGCOMM '18.\hskip 1em plus 0.5em minus 0.4em\relax New York, NY, USA:
  Association for Computing Machinery, 2018, pp. 313--326. [Online]. Available:
  \url{https://doi.org/10.1145/3230543.3230557}
\BIBentrySTDinterwordspacing

\bibitem{Eunomia}
\BIBentryALTinterwordspacing
S.~Mahmood, J.~Lu, and S.~Ghorbani, ``Orderly management of packets in rdma by
  eunomia,'' in \emph{Proceedings of the 9th Asia-Pacific Workshop on
  Networking}, ser. APNET '25.\hskip 1em plus 0.5em minus 0.4em\relax New York,
  NY, USA: Association for Computing Machinery, 2025, pp. 17--23. [Online].
  Available: \url{https://doi.org/10.1145/3735358.3735392}
\BIBentrySTDinterwordspacing

\bibitem{kv-consistency}
E.~Anderson, X.~Li, M.~Shah, J.~Tucek, and J.~Wylie, ``What consistency does
  your key-value store actually provide?'' \emph{HP Laboratories Technical
  Report}, 01 2010.

\bibitem{NetChain}
\BIBentryALTinterwordspacing
X.~Jin, X.~Li, H.~Zhang, N.~Foster, J.~Lee, R.~Soul{\'e}, C.~Kim, and
  I.~Stoica, ``{NetChain}: Scale-free {sub-RTT} coordination,'' in \emph{15th
  USENIX Symposium on Networked Systems Design and Implementation (NSDI
  18)}.\hskip 1em plus 0.5em minus 0.4em\relax Renton, WA: USENIX Association,
  Apr. 2018, pp. 35--49. [Online]. Available:
  \url{https://www.usenix.org/conference/nsdi18/presentation/jin}
\BIBentrySTDinterwordspacing

\bibitem{P6}
A.~Shukla, K.~Hudemann, Z.~Vági, L.~Hügerich, G.~Smaragdakis, A.~Hecker,
  S.~Schmid, and A.~Feldmann, ``Fix with {P6}: Verifying programmable switches
  at runtime,'' in \emph{IEEE INFOCOM 2021 - IEEE Conference on Computer
  Communications}, 2021, pp. 1--10.

\bibitem{Chimera}
J.~Kim, D.~J. Tian, and B.~E. Ujcich, ``Chimera: Fuzzing p4 network
  infrastructure for multi-plane bug detection and vulnerability discovery,''
  in \emph{2025 IEEE Symposium on Security and Privacy (SP)}, 2025, pp.
  3088--3106.

\bibitem{MC-TCP}
M.~Musuvathi, D.~R. Engler \emph{et~al.}, ``Model checking large network
  protocol implementations,'' in \emph{NSDI}, vol.~4, 2004, pp. 12--12.

\bibitem{Plankton}
\BIBentryALTinterwordspacing
S.~Prabhu, K.~Y. Chou, A.~Kheradmand, B.~Godfrey, and M.~Caesar, ``{Plankton}:
  Scalable network configuration verification through model checking,'' in
  \emph{17th USENIX Symposium on Networked Systems Design and Implementation
  (NSDI 20)}.\hskip 1em plus 0.5em minus 0.4em\relax Santa Clara, CA: USENIX
  Association, Feb. 2020, pp. 953--967. [Online]. Available:
  \url{https://www.usenix.org/conference/nsdi20/presentation/prabhu}
\BIBentrySTDinterwordspacing

\bibitem{fairness}
Y.~Kesten, A.~Pnueli, L.-O. Raviv, and E.~Shahar, ``Model checking with strong
  fairness,'' \emph{Formal Methods in System Design}, vol.~28, no.~1, pp.
  57--84, 2006.

\end{thebibliography}

\clearpage \newpage
\appendices

\section{Specification Language Grammar}
\label{sec:grammar}

\begin{figure}[!ht]
\centering
\small
\begingroup
\renewcommand{\c}[1]{`\text{#1'}}
\begin{align*}
spec     &::= && block^* \\
block    &::= && \mathbf{configuration}\ \c\{\ config^*\ \c\}\\
&\ \ |        && \mathbf{topology}\ \c\{\ topo^*\ \c\} \\
&\ \ |        && \mathbf{protocol}\ \c\{\ protocol^*\ \c\} \\
&\ \ |        && \mathbf{property}\ \c\{\ property^*\ \c\} \\
config   &::= && assign \\
assign   &::= && var\ \c=\ value \\
topo     &::= && \mathbf{nodetype}\ type^* \\
&\ \ |        && type\ node^* \\
&\ \ |        && \mathbf{link}\ node^*\ `--\text{'}\ node^* \\
&\ \ |        && \mathbf{route} \c(\ src^*\ \c)\ \c\{\ route\_entry^*\ \c\}\ \\
route\_entry
         &::= && dst^*\ \c:\ next\_hop \\
protocol &::= && \mathbf{var} \c(\ type\ \c)\ assign^* \\
&\ \ |        && \mathbf{thread} \c(\ type\ \c)\ name\ \c\{\ stmt^*\ \c\}\ \\
stmt     &::= && breakpoint\ \c: \\
&\ \ |        && exp \\
&\ \ |        && assign \\
&\ \ |        && \mathbf{temp}\ assign^* \\
&\ \ |        && \mathbf{if}\ \c(\ exp\ \c)\ \c\{\ stmt^*\ \c\}\ \\
&\ \ |        && \mathbf{while}\ \c(\ exp\ \c)\ \c\{\ stmt^*\ \c\}\ \\
&\ \ |        && \dots \\
exp      &::= && \mathbf{forall}\ var\ \mathbf{in}\ set\ \c{:}\ exp \\
&\ \ |        && \mathbf{exists}\ var\ \mathbf{in}\ set\ \c{:}\ exp \\
&\ \ |        && exp\ op\ exp \\
&\ \ |        && func\ \c(\ param^*\ \c) \\
&\ \ |        && \dots \\
func     &::= && \mathbf{Send\ |\ Multicast\ |\ Receive\ |\ Wait} \\
&\ \ |        && \mathbf{Exit\ |\ Assert\ |\ \dots} \\
property &::= && name\ \c=\ ltl \\
ltl      &::= && exp\ |\ \c{[]}\ exp\ |\ \c{\textless\textgreater}\ exp\ |\ \dots \\
\end{align*}
\endgroup
\caption{The grammar of \self's specification language, simplified to highlight the core structure.}
\label{fig:grammar}
\end{figure}

Figure~\ref{fig:grammar} presents the core structure of the language. The complete syntax and semantics are available in the open-source implementation.

\section{Example Thread}
\label{sec:example-thread}

\begin{figure}[!ht]
\centering

\begingroup
\definecolor{comment}{rgb}{0, 0.5, 0}
\definecolor{type}{rgb}{0.17, 0.57, 0.69}
\definecolor{number}{rgb}{0, 0.5, 0}
\definecolor{string}{rgb}{0.75, 0.08, 0.08}
\definecolor{func}{rgb}{0.54, 0.17, 0.89}

\lstset{
    backgroundcolor=\color{white},   %
    frame=single,                    %
    rulecolor=\color{black},         %
    basicstyle=\ttfamily\small,      %
    numbers=left,                    %
    numberstyle=\tiny\color{gray},   %
    breaklines=true,                 %
    commentstyle=\color{comment},    %
    comment=[l]{//},
    stringstyle=\color{comment},
    string=[b]",
    showstringspaces=false,
    keywordstyle=[1]\color{blue},    %
    keywordstyle=[2]\color{type},
    keywordstyle=[3]\color{func},
    keywords=[1]{thread, with, temp},
    morekeywords=[2]{Client, Switch, Controller, Server},
    morekeywords=[3]{Send, Receive, Assert, Exit},
    literate=
        {c1}{c1}2
        {c2}{c2}2
        {c3}{c3}3
        {0}{{{\color{number}0}}}1
        {1}{{{\color{number}1}}}1
        {2}{{{\color{number}2}}}1
        {3}{{{\color{number}3}}}1
        {4}{{{\color{number}4}}}1
        {5}{{{\color{number}5}}}1
        {6}{{{\color{number}6}}}1
        {7}{{{\color{number}7}}}1
        {8}{{{\color{number}8}}}1
        {9}{{{\color{number}9}}}1
}

\begin{lstlisting}
thread(Client) LockOp {
Acquire:
  temp acquire_request = ...;
  Send(acquire_request);
Release:
  temp pkt = Receive();
  Assert(pkt.type == GRANT, "unexpected packet type");
  temp release_request = ...;
  Send(release_request);
End:
  temp pkt = Receive();
  Assert(pkt.type == ACK_OF_RELEASE,
      "unexpected packet type");
  Exit();
}
\end{lstlisting}
\endgroup

\caption{An example thread.}
\label{fig:example-procedure}
\end{figure}

Figure~\ref{fig:example-procedure} shows an example thread, in which the client acquires a lock and then releases it.

\section{Compilation and Model Checking}
\label{sec:compilation}

\sysname uses TLA+ as its model checker. The \sysname compiler converts the user-defined INC system model to the state transition representation accepted by TLA+. Most of \sysname's language elements concerning computation can be mapped directly to those in TLA+, including variable operations and control flows.

In model checking, each thread operates as an independent concurrent unit. Variable changes between two adjacent breakpoints constitute a single transition. Since a transition is atomic and its intermediate states are invisible to other threads, modifying the same variable twice within a transition is either unnecessary or indicates a missing breakpoint. If the intermediate value is only needed for local computation, the user replaces the first assignment with a \texttt{temp} statement, which defines a temporary value that is not part of the system state. If the intermediate value should be observable, the user inserts a breakpoint between the two assignments so that each value is exposed as a distinct global state. \self maintains a dictionary, mapping each thread to its execution point (breakpoint), updated in each transition. \self ensures strong fairness~\cite{fairness}: any thread not indefinitely blocked eventually executes, which makes sense in reality. A thread terminates when it either finishes its code or explicitly calls \texttt{Exit}; in the latter case, all threads within the same node also terminate. A terminated thread is permanently blocked. The system terminates when every thread terminates.

\self provides users with global variables to maintain cross-node state. Updates to these variables are atomic within a transition. They are used only for verification and do not represent shared storage in the verified system.

Specific network abstractions in \sysname are realized with additional states. \self maintains a dictionary that maps each node to a sequence of packets, denoting its receive buffer. \self also maintains a routing table, which is a dictionary that maps source-destination pairs to the next-hop nodes, consulted when sending packets. The table is partially specified by the user and finalized by the \self compiler according to the network topology. The compiler throws an error if multiple routing paths exist but not specified by the user.

\texttt{Receive} blocks until the node's receive buffer is non-empty and retrieves and removes the first element. Receive buffers are unbounded by default; users can impose a capacity limit in the specification when buffer overflow is relevant. The behavior of \texttt{Send} varies by network environment: in a reliable network, it appends the packet to the tail of the next hop's receive buffer; in a lossy network, it might drop the packet; in an out-of-order network, it can insert the packet at any position in the next hop's receive buffer.

The reporter operates inversely to the compiler. It takes the violation trace produced by the model checker, identifies the special variables maintained by \self, reconstructs the network elements, and presents them to the user. Based on the violation trace, users can gain insights into potential risks and refine the protocol design.

\section{Cache Consistency Definition and Checking}
\label{sec:cache-consistency}

\begin{figure}[!ht]
\centering
\includegraphics[width=0.47\textwidth]{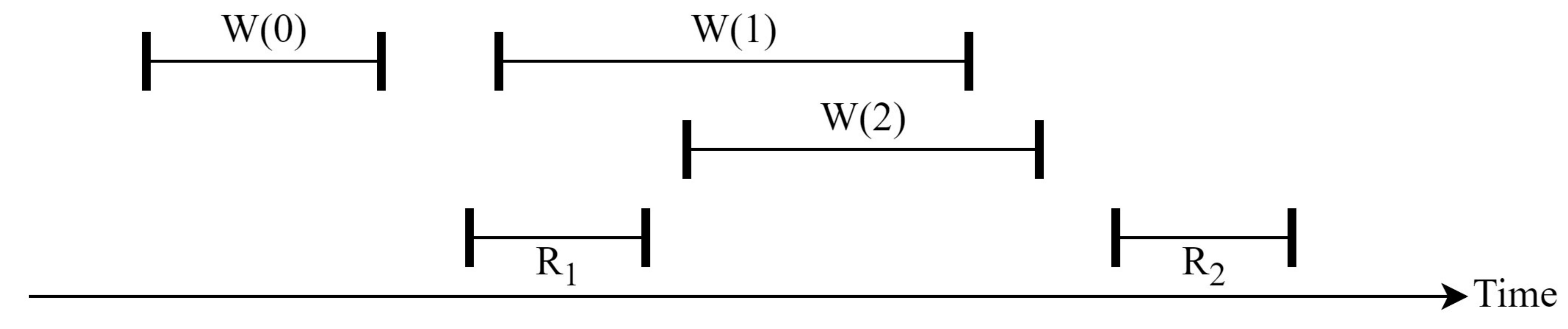}
\caption{An example of cache consistency. All operations perform on the same item. \textnormal{W}($n$) means writing its value to $n$. $\mathrm{R_1,R_2}$ are two reads.}
\label{fig:coherence-example}
\end{figure}

A key-value store system is said to be consistent if each individual item within it is consistent. For a specific item, an operation on it is either a read or a write, associated with a start time and an end time. Two operations are considered concurrent if their time intervals overlap; otherwise, their order can be determined. An item is said to be consistent if every read operation that is not concurrent with any write returns one of the values of the most recent writes. A read operation that is concurrent with some write may return any value. Figure~\ref{fig:coherence-example} shows an example of cache consistency. $\mathrm{R_1}$ may return any value as it is concurrent with W(1). $\mathrm{R_2}$ must return 1 or 2 as W(1) and (2) are the most recent writes.

\begin{algorithm}[tb]

\caption{Cache Consistency Checking}
\label{alg:cache-coherence}
\small

\DontPrintSemicolon
\SetAlgoLined
\SetKwInOut{KwInit}{Init}

\KwIn{An event $e$}
\KwInit{number of ongoing writes $num=0$, ongoing reads $reads=\emptyset$ which are not concurrent with writes, possible latest values $values=\{$initial value$\}$}

\uIf{$e$ \textnormal{is the start of a write}}{
    $num = num + 1$\;
    $reads = \emptyset$\;
    $values = \emptyset$\;
}
\uElseIf{$e$ \textnormal{is the end of a write}}{
    $num = num - 1$\;
    $values = values \cup \{e.value\}$
}
\uElseIf{$e$ \textnormal{is the start of a read}}{
    \If{$num == 0$}{
        $reads = reads \cup \{e.id\}$
    }
}
\Else(\tcc*[f]{end of a read}){
    \If{$e.id\in reads$}{
        \lIf{$e.value \notin values$}{
            report violation
        }
        $reads = reads - \{e.id\}$
    }
}

\end{algorithm}

In \self, an operation starts when the client sends a request and ends when it receives a reply. Each operation is assigned a unique ID. Retransmitted requests are treated as new operations. We propose Algorithm~\ref{alg:cache-coherence} to check cache consistency. It is implemented as a check function, invoked at the start and end of each operation (referred to as an event), terminating the verification as soon as a violation is detected. Users can adopt another definition of consistency by replacing the check function.

\section{Verification Result of Tensor Aggregation Systems}
\label{sec:tensor-aggregation-details}

\parab{Protocol-specific models.}

SwitchML maintains an aggregator array of size $2S$ on the switch. A request is mapped to an aggregator through a modulo operation. Clients initially send the first $S$ requests. Upon receiving the result for request $t$, the client sends request $t+S$, continuing until there are no more requests. Consequently, if a specific position in the aggregator is occupied by request $n$ and later reoccupied by $n+2S$, it indicates that request $n+S$ has been completed, which in turn implies that all clients have received the result for request $n$, allowing its information to be safely overwritten. SwitchML does not involve the PS.

ATP targets multi-tenant scenarios. To maximize spatial utilization under multi-task parallelism, ATP maintains a single aggregator array on the switch. Each request of a task is hashed to a specific position within the array. In cases of hash collisions, the aggregation falls back to the PS. When the switch receives a retransmitted request, it releases the corresponding aggregator and falls back to the PS to prevent the aggregator from being permanently occupied. As discussed in Section~\ref{sec:optimization:item}, we only focus on a specific task.

NetReduce is dedicated to in-network aggregation compatible with Remote Direct Memory Access over Converged Ethernet (RoCE), a widely used data-center transport for low-latency remote memory access. RoCE splits a message into multiple packets, but only the first packet carries the aggregation header, which is necessary for addressing. NetReduce maintains a Connection Lookup Table (CLT) on the switch, mapping connections to aggregation headers. When the first packet of a message arrives, the switch records the aggregation header in the CLT. When subsequent packets arrive, the switch queries the CLT to obtain the header. NetReduce ensures that a packet belongs to the same message as the one recorded in CLT by comparing the difference in packet sequence number (PSN). If they do not belong to the same message, the packet is discarded. NetReduce does not involve the PS.

\parab{Property violations.}

$\bullet$ \textit{Computational correctness violation in SwitchML.} Swi\-tch\-ML does not maintain PSN in aggregators. Due to network delays, the reply to request $t$ may not arrive before a timeout, causing the client to retransmit request $t$. When the delayed reply finally arrives, the client proceeds to send request $t+S$. In an out-of-order network, the retransmitted request $t$ may arrive at the switch after the completion of request $t+S$, and is involved in the aggregation of request $t+2S$, which leads to incorrect results.

$\bullet$ \textit{Memory leak freedom violation in ATP.} ATP frees an aggregator under two conditions: (1) the aggregation is completed\footnote{ATP sends the aggregation result to the PS for durability, and frees the aggregator when receiving an ACK from the PS.}, and (2) a retransmitted request is received. However, in an out-of-order network, an original request may be trapped in the network, and reaches the switch after the aggregation has been completed with its retransmission. If the request takes up an aggregator, it cannot be normally freed.

$\bullet$ \textit{Terminality violation in NetReduce.} Assume that RoCE splits each message into $n$ packets, and all requests are sequentially numbered as $1.1,1.2,\dots,1.n,2.1,\dots$, with the window size at least two messages. The following trace is possible. Requests $1.1$ from all clients arrive at the switch and update the CLT, completing the aggregation. Some request in $1.2,\cdots,1.n$ from a client is lost, yet his request $2.1$ reaches the switch and updates the CLT before retransmission of the lost request. At this point, the CLT does not contain the aggregation header for message 1. Since the aggregation has been completed, the client will never retransmit request $1.1$. As a result, the lost request will keep being retransmitted by the client and dropped by the switch, which is a livelock.

\section{Verification Result of Lock Management Systems}

\label{sec:lock-management-results}

\parab{System model.} We model a lock management system using a star topology, where a switch connects multiple clients and a lock server. As discussed in Section~\ref{sec:optimization:item}, we only focus on a specific lock. Clients send requests to the lock server to acquire or release that lock. INC protocols allow the switch to maintain lock metadata and serve lock requests. We choose NetLock~\cite{NetLock} and \fisslock~\cite{FISSLOCK} as representative protocols.

NetLock maintains a queue for each lock in the switch. An element in the queue is a tuple of the lock mode, transaction ID, and client IP. The first few elements represent the current holders, while the remaining represent the waiters. Due to the limitations of programmable switches, only one position in the queue can be accessed while processing a packet. When receiving a lock acquisition request, the switch enqueues it. If the queue was previously empty (no holder), or elements in the queue and the request are all in shared mode, the request is directly granted. When receiving a release request, the switch dequeues the head element and resubmits the request to the beginning of the processing pipeline, granting the new head if not granted yet. Because the switch can only dequeue the head of the queue, it does not check the transaction ID when releasing locks. NetLock argues that this design does not affect the correctness, because only one transaction can hold an exclusive lock, and the operations of releasing shared locks are commutative.

\fisslock decouples lock management into grant decision and participant maintenance. The switch is only responsible for grant decisions, maintaining the lock mode (free, exclusive, or shared), and the holder's machine ID for each lock. The lock agent, which resides on the lock holder, is responsible for participant maintenance, i.e., the holder and the waiters. When the holder changes, the agent migrates. When receiving a lock acquisition request, the switch decides whether to grant the client based on lock mode. If the lock is free, the switch replies to the client with a grant packet and asks it to establish the lock agent. If both the lock and the acquisition request are shared, the switch replies to the client with a grant packet and forwards the request to the lock agent. In other cases where grant is not immediately possible, the switch simply forwards the request to the lock agent. When receiving a lock release request, the switch forwards it to the lock agent. When there is no holder, the lock agent migrates itself to the next waiter, or notifies the switch to free the lock if no waiter exists. If the client and the agent are on the same machine, acquisition and release will be handled locally without passing through the switch.

\fisslock discusses anomalies caused by network exceptions and introduces patch mechanisms. To address packet loss, \fisslock requires each client-initiated packet to be acknowledged, and retransmitted after a timeout. To prevent metadata from being modified twice, the switch maintains a sequence number for each server to track the number of processed packets, and only updates the metadata if the incoming packet's sequence number is higher than the on-switch number. One exception is the grant packet, which is initiated by the switch. When a lock acquisition times out, \fisslock requires the client to release the lock and retry the acquisition. To address packet out-of-order (e.g., release before acquisition), \fisslock requires the agent to send unprocessable packets back to the switch. This additional routing corrects packet order.

\fisslock discusses another anomaly caused by a particular event interleaving. When a shared lock is released, the agent grants the lock to the first waiter, who attempts to acquire the lock exclusively, and migrates itself to that machine. Meanwhile, the switch grants the same shared lock to another client, violating lock exclusion. To address this anomaly, \fisslock introduces the incarnation mechanism. Both the switch and the agent maintain a per-lock incarnation, which is incremented upon receiving a shared acquisition. The agent embeds its incarnation in grant and free packets it sends. When the switch receives such a packet, it compares the embedded incarnation with its own. If they match, the switch resets the incarnation and handles the request as usual. Otherwise, it rejects the packet, and the agent returns to its original machine, continuing to handle lock requests.

\parab{Correctness properties.} We specify the following correctness properties with \sysname.

$\bullet$ \textit{Terminality}, the same as described in Section~\ref{ssec:tensor-aggregation-results}.

$\bullet$ \textit{Lock exclusion}, which requires that if a lock is held by one client exclusively, it cannot be held by other clients. This property is verified by checking state variables in the system.

\parab{Property violations.}

$\bullet$ \textit{Lock exclusion violation in NetLock.} NetLock does not provide a detailed discussion on retransmission. Without special handling, the following trace is possible. At some moment, the lock is held in shared mode by client 1 and client 2. Client 1 initiates a lock release request, but it is delayed in the network, triggering a timeout and resulting in a retransmission of the release request. Since the switch does not check transaction ID, it processes both release requests and dequeues two elements. If client 3 initiates an exclusive acquisition request at this point, the switch grants it as the queue is empty. This violates lock exclusion: client 3 holds the lock exclusively, while client 2 still holds it in shared mode.

\begin{figure}[!ht]
\centering
\includegraphics[height=0.4\textheight]{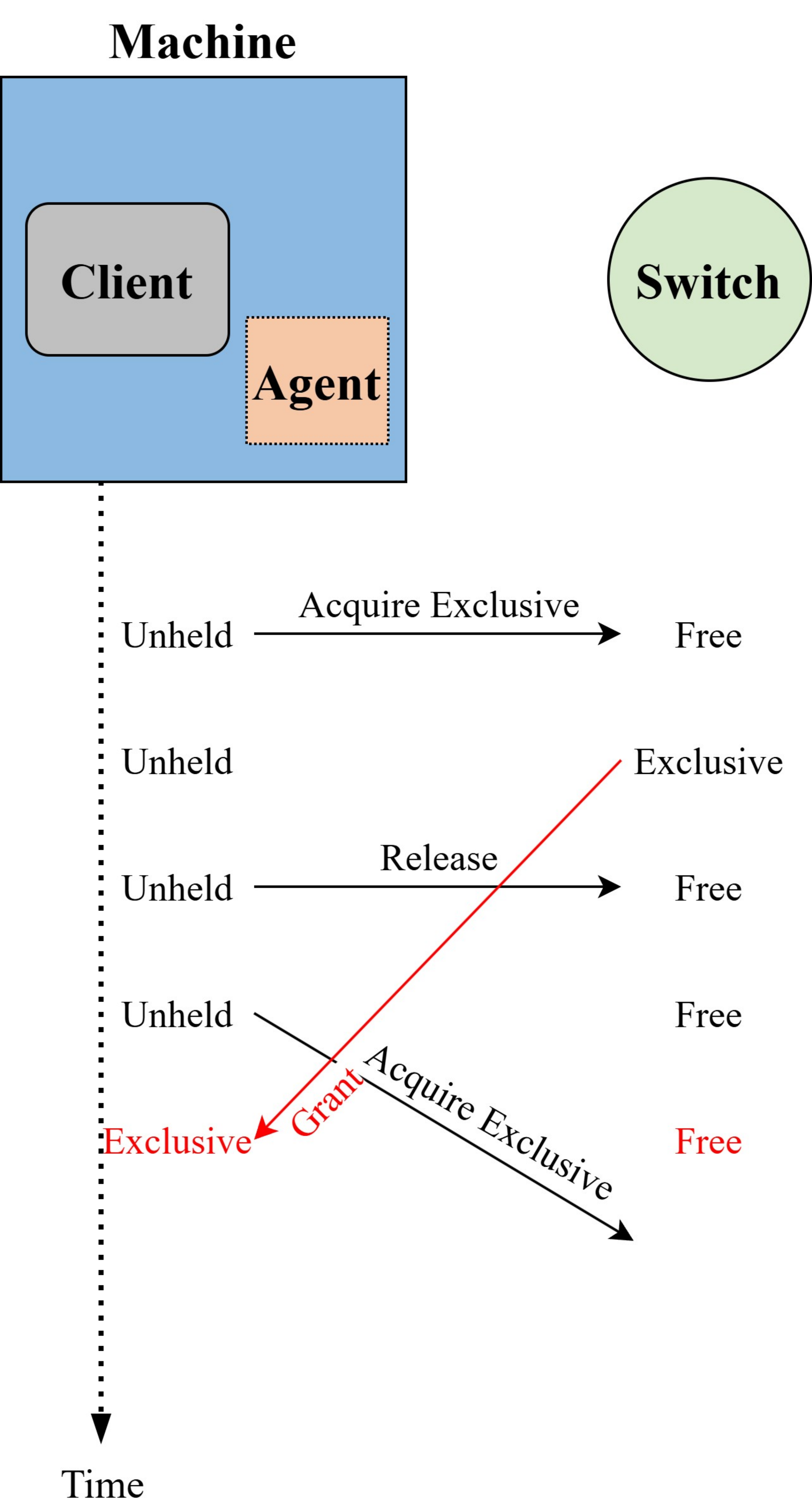}
\caption{An execution trace of lock exclusion violation in \fisslock. All operations perform on the same lock, which is initially free and cached in the switch. Important events are marked in red.}
\label{fig:fisslock-ds-trace}
\end{figure}

$\bullet$ \textit{Lock exclusion violations in FissLock.} Figure~\ref{fig:fisslock-ds-trace} illustrates an execution trace of lock exclusion violation in \fisslock. The lock is initially free. The client initiates a lock acquisition request, and the switch replies with a grant packet. However, it is delayed in the network, leading to a timeout that triggers the release and re-acquisition of the lock. Later, the client receives the delayed grant and assumes it holds the lock, while the switch, having handled the release request, considers the lock free.
At this point, if a client on another machine attempts to acquire the lock at this point, the switch will directly grant it. This will violate lock exclusion if one of the two clients holds the lock exclusively.

\section{Possible Fixes to Identified Violations}
\label{sec:fixes}

SwitchML works correctly under reliable networks. The \textit{computational correctness violation in SwitchML under out-of-order networks} can be fixed by associating a PSN with each aggregator. Packets carrying a different PSN are excluded from the aggregation process of an aggregator.

ATP works correctly under reliable networks. The \textit{memory leak freedom violation in ATP under out-of-order networks} can be fixed by associating a timestamp with each aggregator and enabling the system to preempt aggregators that have timed out in favor of new requests. This is actually the idea of DSA~\cite{DSA}.

NetReduce works correctly under reliable networks. The \textit{terminality violation in NetReduce under lossy networks} can be fixed by incorporating PSN into the key of CLT, so that CLT entries are uniquely identified and will not be overwritten unintentionally.

The \textit{cache consistency violation in NetCache} can be fixed by requiring the server, upon receiving a write request, to first confirm that the switch cache has been updated before replying to the client. Yet, this may harm system performance.

The \textit{terminality violation in NetCache} can be fixed by requiring the switch to respond with a special message when it is asked to update a non-existent item. Other concurrent events should be handled similarly.

The \textit{cache consistency violation in FarReach} can be fixed by requiring the switch to monitor all replies from the server and update its cache whenever the reply carries a newer value.

The \textit{lock exclusion violation in NetLock} can be fixed by requiring the switch to check the transaction ID and dequeue the correct element upon receiving a release request. While this solution is conceptually straightforward, it is challenging to implement within the constraints of a programmable switch.

It is difficult to provide a simple fix to the \textit{lock exclusion violations in FissLock}. As \fisslock migrates the lock agent across the entire network, it is challenging to maintain the consistency between the lock agent, the switch, and the actual state of the system. While only two violation traces are presented, additional ones have been observed. Although distributed consensus algorithms are applicable, they may incur significant performance overhead.

\end{document}